%% file: main2.tex
\pdfoutput=1
\documentclass[%
 reprint,amsmath,amssymb,aps,prb]{revtex4-2}
\usepackage{dcolumn}
\usepackage{comment}
\usepackage{float}
\usepackage{color}
\usepackage{wrapfig}
\usepackage{subcaption}
\usepackage{url}
\usepackage[pdftex]{graphicx}
\usepackage{enumerate}
\usepackage{amsmath}
\usepackage{bm}
\usepackage{bigints}
\usepackage{braket}
\usepackage{autobreak}
\usepackage[version=3]{mhchem}
\usepackage{mathrsfs}
\usepackage{here}
\usepackage{mathtools}
\usepackage{physics}
\renewcommand{\emph}[1]{\textnormal{#1}}
\begin{document}
\preprint{APS/123-QED}
\title{Dynamics of Topological Defects in Type-II Superconductors \\
under Gradients of Temperature/Spin Density}

\author{Takuma Kanakubo}\email{kanakubo@vortex.c.u-tokyo.ac.jp}
\affiliation{Department of Physics, The University of Tokyo, Bunkyo-ku, Tokyo, 113-0033, Japan}

\author{Hiroto Adachi}
\affiliation{Research Institute for Interdisciplinary Science, Okayama University, Okayama 700-8530, Japan}


\author{Masanori Ichioka}
\affiliation{Research Institute for Interdisciplinary Science, Okayama University, Okayama 700-8530, Japan} 

\author{Yusuke Kato}\email{yusuke@phys.c.u-tokyo.ac.jp}
\affiliation{Department of Basic Science, The University of Tokyo, Meguro, Tokyo 153-8902, Japan}
\affiliation{Department of Physics, The University of Tokyo, Bunkyo-ku, Tokyo, 113-0033, Japan}

\begin{abstract}
We theoretically investigate the motion of a domain wall and a vortex
in type-II superconductors driven by inhomogeneities of temperature or
spin accumulation. The model consists of the time-dependent
Ginzburg-Landau equation and the thermal or spin diffusion equation,
whose transport coefficients, such as the thermal and spin
conductivities and the spin relaxation time, depend on the order
parameter and interpolate between their values in the superconducting
and normal states. Numerical and analytical calculations indicate that
the domain wall moves toward the higher-temperature region or the region
with larger spin accumulation, where the order parameter is suppressed.
We also derive analytical expressions for the vortex velocity and confirm
the predicted direction of vortex motion by numerical simulations. The
dynamics of these topological defects can be understood as processes
that reduce the loss of condensation energy. We also analyze the driving
force, viscous force, thermal force, and force due to the spin
accumulation gradient on the basis of momentum balance relations.
\end{abstract}

\maketitle

\input{body2}

\bibliography{ref}
\bibliographystyle{apsrev4-2}  

\appendix

\clearpage

\onecolumngrid 
\section*{Supplemental Material}
\renewcommand{\theequation}{S\arabic{equation}}
\setcounter{equation}{0} 

\input{SM2}

\end{document}

%% file: body2.tex
\section{\label{sec:level1}Introduction}
Quantum vortices in type II superconductors have been a central focus of research. Quantum vortices are the singularities in the phase of the order parameter and behave stably as topological defects or emergent particles against the background of the superconducting condensate. They play a vital role in the Berezinskii-Kosterlitz-Thouless (BKT) transition \cite{berezinskii1971destruction,Berezinsky1972rfj,kosteritzthouless1973,halperin1979}, a characteristic transition mechanism in two-dimensional systems. 

Type II superconductors in magnetic fields exhibit several phases, such as the vortex lattice, vortex glass, and vortex liquid \cite{blatter,nattermann}, collectively referred to as vortex matter. 

Exploring an efficient method for controlling vortex motion in superconductors is significant from an application perspective. In addition to driving by transport electric currents \cite{GORTER1962,DeGennes1964,strnad1964dissipative,kim1965flux,Schmid66,GK1971,GK1975,Dorsey92,Chen98,narayan2003driving,chen2010numerical,KC,Sugai}, driving vortices via heat flow has been a major approach and has been extensively studied since the 1960s \cite{Maki65,Stephen66,OtterSolomon66,HUEBENER1967.947,Ohta1967,yntema1967thermodynamics,solomonotter1967thermomagnetic,SOLOMON1968293,Maki1969,Huebener69,Lange1974,
Kopnin1975,
samoilov1992vortex,huebener2001}. 

Such research elucidates the dynamics of superconductors in magnetic
fields while also providing broader insights into the dynamics of
topological defects in various ordered states.

Typical examples include domain walls and skyrmions in magnets\cite{bogdanov1989,NagaosaTokura2013,Kong2013,Lin2013,
burelbach2018unified,yu2021real}
and dislocations in smectic liquid crystals\cite{chaikin1995,Pleiner1986,Blanc2004}.
A common theme in these systems is the motion of topological defects
driven by external fields or gradients. Among such problems, thermally
driven motion is particularly relevant to the present work. In
thermally driven cases, both the direction of motion and the underlying
mechanism can depend sensitively on the specific driving mechanism: the
texture or defect may move either along or against the temperature gradient.
\color{black}
Vortices in superconductors are no exception to this complexity. Recently, as described below, a controversy exists on the interplay between heat flow and vortices. Furthermore, several groups \cite{KimSeKwon2018,vargunin2019flux,Taira2018,adachi2024timedependent} have begun to investigate the interplay between spin current and vortices as a new route for controlling vortex motion and developing spintronic devices.

In this paper, we address the motion of topological defects, a domain wall and a vortex, under heat flow and spin current. 
What we call a domain wall in the present work is a boundary-condition-imposed one-dimensional configuration in which the
superconducting order parameter varies along a single spatial direction
and changes sign between the two boundaries~\cite{richard2016heat} (see Figure~\ref{figsystem}).
We use this configuration as a simplified topological-defect prototype.
A domain wall of this type is physically realized in the
Fulde-Ferrell-Larkin-Ovchinnikov (FFLO) state~\cite{ff,LO}, in which the
order parameter undergoes periodic spatial modulations forming a
sequential array of domain walls, as confirmed
experimentally~\cite{mayaffre2014evidence}. The characteristic length scale of this variation is the coherence length, just as in the case of vortices. While vortices can form by applying a magnetic field exceeding the critical field, the existence of the domain wall is ensured by the boundary conditions imposed on the order parameter (see Eqs.~\eqref{bc of OP}). A related use of such a simplified one-dimensional setting can be found in Ref.~\cite{stone1996}, where a domain wall problem was used as a concise case study for discussing the dynamics of Andreev bound states associated with vortices.

In the present study, we use the DW model as a tractable conceptual prototype to study the basic mechanism of gradient-driven defect motion in the simplest setting before turning to the vortex problem. As shown below, the velocity expressions obtained for the DW and for the vortex have closely analogous structures under both temperature and spin-density gradients. This correspondence motivates the use of the DW analysis as a preliminary step, while the analysis for a single vortex in a magnetic field is subsequently carried out in Secs.~IV and V.

\color{black}

In the following two paragraphs, we provide an overview of the current understanding of vortex motion under heat flow and spin current. Subsequently, we will present our aim in this paper more explicitly. 

\vskip\baselineskip
\paragraph{heat flow}
Studies on the motion of vortices under temperature gradients date back
to the 1960s~\cite{Maki65,Stephen66,OtterSolomon66,HUEBENER1967.947,Ohta1967,yntema1967thermodynamics,solomonotter1967thermomagnetic,SOLOMON1968293,Maki1969,Huebener69,Lange1974,
Kopnin1975,
samoilov1992vortex, freimuth2000spectral, huebener2001, chen2021manipulation, niedzielski2026thermal}. 

Stephen derived, in his pioneering theoretical study~\cite{Stephen66}, an expression for the effective thermal force on a vortex line per unit length
\begin{equation} \label{Fthstephen}
  \bm{F}_{\textrm{th}} = -\frac{\phi_0 S}{|B|} \grad T,
\end{equation}
on the basis of thermodynamic considerations.
Here, $S$ and $B$ denote the entropy per unit volume and the magnetic induction, respectively. The quantity $\phi_0=h/2e$ is the superconducting flux quantum, with
$h$ the Planck constant and $e(>0)$ the elementary charge. Equation~\eqref{Fthstephen} shows that the thermal force is antiparallel to the gradient of temperature $T$, and thus, Stephen stated that the vortex is driven from the hotter to the colder region. Yntema wrote an equation similar to Eq.~\eqref{Fthstephen} in~\cite{yntema1967thermodynamics}. 

Thermoelectric effects, namely the Nernst and Ettingshausen effects, in vortex systems were also investigated experimentally in the same
period~\cite{OtterSolomon66,solomonotter1967thermomagnetic,HUEBENER1967.947,SOLOMON1968293}. These early studies assumed that vortices move toward the lower-temperature region under a temperature gradient.

In Stephen's paper, the coefficient $-\frac{\phi_0 S}{B}$, which appears in Eq.~\eqref{Fthstephen}, is explained as ``the entropy of unit length of a vortex line''. However, it is not immediately evident how this can be related to what is currently known as ``the transport entropy'', as pointed out in~\cite{solomonotter1967thermomagnetic}.

The interpretation of the transport entropy was later revisited by
Sergeev \textit{et al.}~\cite{sergeev2008heat,Sergeev2010}.

In their thermodynamic argument, the force contribution associated with
the electromagnetic energy of the circulating supercurrents is canceled
by the Lorentz force generated by temperature-gradient-induced
magnetization currents. They therefore argued that the circulating
supercurrents do not carry entropy or heat, and that the transport
entropy should be attributed to the ordinary thermodynamic entropy
localized in the vortex core.

\color{black}
Studies of related systems continue to the present day in the context of thermoelectric effects in vortex liquids~\cite{Behnia_2023}.

Several studies have suggested that vortices can move toward, or become localized in, the higher-temperature region~\cite{Veshchunov2016,Duarte2019,DETOLEDO2021,niedzielski2026thermal}.

Veshchunov \textit{et al.} conducted experiments to thermally control quantum vortices ~\cite{Veshchunov2016}. Their experiment revealed that the local temperature gradient generated by the laser drives the vortices and attracts them toward the laser spot (i.e., the local high-temperature region), thereby facilitating their easy positioning. Their result suggests that driving vortices by temperature gradients can be achieved with high speed and precision~\cite{Veshchunov2016}. The authors of Ref.~\cite{Veshchunov2016}
derived the thermal force on the isolated vortex in the presence of the thermal gradient on the basis of the energetics in the London equation.   
Duarte \textit{et al.}~\cite{Duarte2019} and de Toledo \textit{et al.}~\cite{DETOLEDO2021} numerically solved the time-dependent Ginzburg-
Landau (TDGL) equation~\cite{Schmid66,gorkoveliashberg} under a uniform temperature gradient and reported that the vortices begin to form in the hotter region.

These situations imply that the following two issues are important.
\begin{enumerate}
  \item Previous studies have not reached a consensus on the direction of motion of superconducting vortices under a temperature gradient.
  \item 
The origin of the thermal force should be clarified on a firmer basis.
\end{enumerate}
A possible way to address the second issue is to define the force on an
isolated vortex in terms of the stress tensor, as stated in a review
article (see p.~1130 of Ref.~\cite{Kimstephen}), or equivalently, based
on the momentum balance relation.
\ \\
\paragraph{spin current}
Another possible way to drive vortices is the use of spin currents. In recent years,  spintronics in superconductors has attracted many researchers' attention~\cite{Takahashi1999, Yang2010,Taira2018,KimSeKwon2018,umeda2018spin,vargunin2019flux,Vargas2020,taira2021spin,sharma2023spin,adachi2024timedependent}. One of the most intriguing phenomena is the vortex spin Hall effect (vortex SHE)\cite{KimSeKwon2018, vargunin2019flux,taira2021spin,adachi2024timedependent}, in which spin currents are generated by spin accumulation in the vortex and carried along with the vortex motion. Furthermore, the inverse vortex spin Hall effect (inverse vortex SHE \cite{taira2021spin,adachi2024timedependent}) has recently been observed in experiments~\cite{umeda2018spin,sharma2023spin}, attracting significant attention in this field. To understand these new phenomena, developing a theory that describes the interaction between vortices and spin currents is essential, particularly the transport of vortices under spin currents. However, the understanding of the forces on a vortex arising from spin currents is not well established; thus, further research is needed.

This paper aims to describe the motion of a domain wall and a vortex, and to propose force definitions based on the momentum balance relation. Although our final goal is to understand vortex dynamics, we first consider the domain wall as a tractable prototype. Analytical and numerical results for the domain wall provide a simple setting in which the mechanism of gradient-driven defect motion can be examined transparently. 

We then turn to the single-vortex problem and show how an analogous mechanism governs vortex motion.
\textcolor{black}{We use the TDGL equation, which has successfully described the properties of vortices in the flux-flow state. In particular,} we employ the momentum balance relation~\cite{KC,Sugai}, directly derived from the TDGL equation, to discuss the dynamics of topological defects. We also utilize the diffusion equations for the temperature or spin accumulation.

We organize this paper as follows. In Sec.~\ref{sec2gradT}, we introduce the model for the case with a temperature gradient and perform both numerical and analytical calculations. The results of these calculations are consistent with each other. In Sec.~\ref{sec3gradmu}, we address the system with a spin accumulation gradient. In Sec.~\ref{sec4vortex}, we extend the analysis of Sec.~\ref{sec2gradT}
to the dynamics of an isolated vortex under a temperature gradient. In Sec.~\ref{sec_vortex_spin}, we discuss the corresponding vortex motion under a spin-density gradient. In Sec.~\ref{sec3.5discussion}, we discuss possible applications and extensions.

Section~\ref{sec5conc} concludes the present study. 
\section{Motion of domain wall under temperature gradient} \label{sec2gradT}

In this section, we analyze the dynamics of domain wall
under a temperature gradient
in a superconductor with a finite length along the $x$ direction and infinite extensions along the $y$ and $z$ directions. We use the TDGL and thermal diffusion equations, which constitute a boundary-value problem of coupled differential equations. We study the model both numerically and analytically.

\subsection{Model}
\begin{figure*}
\includegraphics[width=110mm]{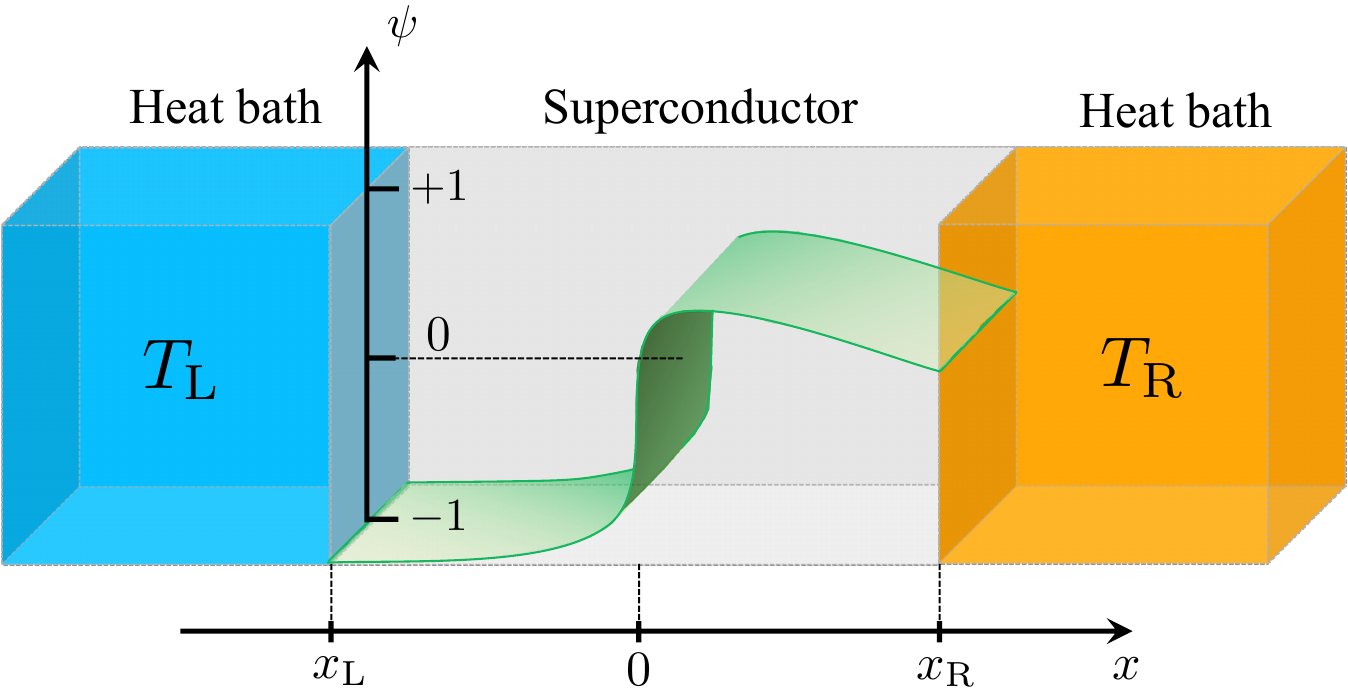}
\captionsetup{justification=raggedright,singlelinecheck=false}
\caption{Schematic picture of the setup we discuss. A sample of the superconductor $(x_{\textrm{L}} \leq x\leq x_{\textrm{R}})$ is placed between heat baths with temperatures $T_{\textrm{L}}$ at $x=x_{\textrm{L}}$ and $T_{\textrm{R}}$ at $x=x_{\textrm{R}}$ respectively. We describe the superconductivity in terms of the dimensionless condensate wavefunction, namely the order parameter $\psi(x,t)$ represented by the vertical axis. We impose the boundary conditions with different signs (see Eqs.~\eqref{boundconds for gradT third} and \eqref{boundconds for gradT last}), and thus we have the domain wall structure of the order parameter in our system.}
\label{figsystem}
\end{figure*}

We consider a superconductor subjected to a temperature gradient within the framework of Ginzburg-Landau theory. We assume that the system is spatially uniform within the $yz$-plane and all physical quantities vary only along the $x$-coordinate. Thus, the problem is reduced to a one-dimensional boundary value problem defined in the region $x_{\textrm{L}} \leq x\leq x_{\textrm{R}}$. For the sake of simplicity, we take $x_{\textrm{L}}=-x_{\textrm{R}}$ in this paper. Figure~\ref{figsystem} shows schematically the system considered in this section, where the domain wall
configuration is imposed by the boundary conditions on the order
parameter.

\subsubsection{TDGL and Thermal diffusion equations}
In the ordinary Ginzburg-Landau theory, the free energy $\mathcal{F}$ per unit area on the $yz$-plane of the three-dimensional system has the form of
\begin{align}
\mathcal{F} &= \int_{x_{\textrm{L}}}^{x_{\textrm{R}}} \text{d}x \  \left[ \frac{\hbar ^2}{2m} \left(\frac{\text{d}\Psi(x)}{\text{d}x}\right)^2 \notag \right. \\& \left. \hspace{0.3cm}  + \alpha_0 \frac{T-T_{\textrm{c}}}{T_{\textrm{c}}}\Psi^2(x) + \frac{\beta}{2} \Psi^4(x) \right].
\end{align} 
Here, $\Psi(x)$ denotes the condensate wavefunction, whose dimension is $(\textrm{volume})^{-1/2}$. We can take $\Psi(x)$ as real due to the absence of the magnetic field. The symbols $m,\hbar$ are the mass of a Cooper pair and the Planck constant divided by 2$\pi$, respectively. The constants $\alpha_0,\beta$ are positive and independent of the temperature and the position. $T_\textrm{c}$ is the critical temperature. Variation of the free energy results in the Ginzburg-Landau equation,
\begin{equation} \label{GLeq}
-\frac{\hbar ^2}{2m} \frac{\text{d}^2\Psi(x)}{\text{d}x^2} + \alpha_0 \frac{T-T_{\textrm{c}}}{T_{\textrm{c}}}\Psi(x) +\beta \Psi^3(x) =0.
\end{equation}
We introduce the time dependence of the order parameter $\Psi(x,t)$ and the temperature $T(x,t)$ to discuss domain wall dynamics. We assume that the time evolution follows the TDGL equation:
\begin{align}\label{eq: TDGL-dimensionful}
  \MoveEqLeft {-\frac{\hbar^2}{2m}\frac{\partial^2 \Psi(x,t)}{\partial x^2} + \alpha_0 \frac{T(x,t)-T_{\textrm{c}}}{T_{\textrm{c}}} \Psi(x,t) + \beta \Psi^3 (x,t)} \notag \\
  & = -\gamma \frac{\partial \Psi(x,t)}{\partial t},
\end{align}
where $\gamma (>0)$ in the RHS is a relaxation coefficient. 

Another equation of motion is the one-dimensional thermal diffusion equation:
\begin{subequations}
\begin{align}  
&C\frac{\partial T(x,t)}{\partial t}+\frac{\partial q(x,t)}{\partial x}=0,\\
&q(x,t)=-\kappa(x,t) \frac{\partial T(x,t)}{\partial x},\label{defofq} 
\end{align}    
\label{originalTDeq}
\end{subequations}
where $C$ is the specific heat per unit volume and $q(x,t)$ is the heat flow density. The symbol $\kappa(x,t)$ denotes the local thermal conductivity, which we assume depends locally on the order parameter $\Psi(x,t)$. We will present the phenomenological expression for 
$\kappa(x,t)$ in the following part in this subsection. 

We impose the boundary conditions
\begin{subequations} \label{bc of T}
    \begin{align}
    \displaystyle
    T\left(x_{\textrm{L}},t\right) & =  T_{\textrm{L}} \label{eq: bc-TL} \\
    \displaystyle
   T\left(x_{\textrm{R}},t\right) & =  T_{\textrm{R}}\label{eq: bc-TR}
    \end{align}
    \label{eq: bc-for-temperature}
\end{subequations}
for the temperature with $T_{\textrm{L}}, \ T_{\textrm{R}} < T_{\textrm{c}}$, and 
\begin{subequations} \label{bc of OP}
    \begin{align}
    \displaystyle
    \Psi\left(x_{\textrm{L}},t\right) & = \Psi_{\textrm{eq}}(T_{\textrm{L}}) \tanh{\frac{x_{\textrm{L}}}{\sqrt{2}\xi}} < 0\label{bc of OPleft} \\
    \displaystyle
    \Psi\left(x_{\textrm{R}},t\right) & = \frac{1}{a}\Psi_{\textrm{eq}}(T_{\textrm{L}}) \tanh{\frac{x_{\textrm{R}}}{\sqrt{2} a \xi}} > 0 \label{bc of OPright}
    \end{align}
    \label{eq: bc-for-OP}
\end{subequations}
for the order parameter. The notation
\begin{equation} \label{PsieqT}
  \Psi_{\textrm{eq}}(T) \equiv \sqrt{\frac{\alpha_0}{\beta}\frac{T_{\textrm{c}}-T}{T_{\textrm{c}}}} 
\end{equation}
is the modulus of the order parameter in bulk in thermal equilibrium at a temperature $T(<T_{\rm c})$. The dimensionless constant
\begin{equation}
  a \equiv \sqrt{\frac{T_{\textrm{c}}-T_{\textrm{L}}}{T_{\textrm{c}}-T_{\textrm{R}}}} = \frac{\Psi_{\textrm{eq}}(T_{\textrm{L}})}{\Psi_{\textrm{eq}}(T_{\textrm{R}})} > 1 
\end{equation}
is the ratio of the modulus of the order parameter in bulk in thermal equilibrium under the temperature \eqref{eq: bc-TL}, \eqref{eq: bc-TR}. The coherence length is defined by
\begin{equation}
   \xi \equiv \sqrt{ \frac{\hbar^2 T_{\textrm{c}}}{2m \alpha_0 (T_{\textrm{c}}-T_{\textrm{L}})}},
\end{equation}
for $T=T_{\textrm{L}}$, and that for $T=T_{\textrm{R}}$ becomes $a\xi$.

We add the factors of the hyperbolic tangent functions to the RHSs of Eqs.~\eqref{bc of OPleft} and~\eqref{bc of OPright} for later convenience. The different signs between  \eqref{bc of OPleft} and \eqref{bc of OPright} force a domain wall to exist somewhere in the region $x_{\textrm{L}} \leq x\leq x_{\textrm{R}}$.

We rewrite the TDGL and the thermal diffusion equations in terms of the dimensionless order parameter 
\begin{equation}
\psi(x,t) \equiv 
\displaystyle \frac{\Psi(x,t)}{\Psi_{\textrm{eq}}(T_{\textrm{L}})}
\end{equation}
and dimensionless temperature 
\begin{equation}
\tau(x,t) \equiv 
\displaystyle \frac{T(x,t)}{T_{\textrm{c}}}.  
\end{equation}
The TDGL equation \eqref{eq: TDGL-dimensionful} becomes
\begin{align} \label{TDGLdimensionless}
\MoveEqLeft{-\xi^2 \frac{\partial^2 \psi(x,t)}{\partial x^2} +\frac{\tau(x,t)-1}{1- \tau_{\textrm{L}}} \psi(x,t) + \psi^3 (x,t)}\notag \\
  & = -\tilde{\gamma} \frac{\partial \psi(x,t)}{\partial t},
\end{align}
where we introduce the relaxation coefficient in Eq.~\eqref{TDGLdimensionless} by 
\begin{equation}
  \tilde{\gamma} \equiv \frac{\gamma T_{\textrm{c}}}{\alpha_0 (T_{\textrm{c}}-T_{\textrm{L}})}.
\end{equation}
We assume $\xi \ll x_{\textrm{R}}$. We note the solution to the equilibrium Ginzburg-Landau equation when we take $x_{\textrm{R}}=\infty$, $x_{\textrm{L}}=-\infty$ and $\tau(x,t)=\tau_{\textrm{L}}$ 
\begin{equation} \label{eqgl0}
  -\xi^2 \frac{\text{d}^2 \psi_0(x)}{\text{d}x^2} -\psi_0(x) + \psi_0^3 (x)=0
\end{equation}
under the boundary conditions $\psi(\pm \infty)=\pm1$ is 
\begin{equation} \label{psi0}
  \psi_0(x) = \tanh{\frac{x}{\sqrt{2}\xi}}.
\end{equation}
Even in the finite system defined in the region $x_{\textrm{L}} \leq x\leq x_{\textrm{R}}$,  Eq.~\eqref{psi0} is the solution to Eq.~\eqref{eqgl0} under the set of the boundary conditions \eqref{eq: bc-for-temperature} and \eqref{eq: bc-for-OP}  when $T_{\rm R}=T_{\rm L}$.
Before rewriting the thermal diffusion equation, we assume that the thermal conductivity depends locally on the modulus of the order parameter in a way similar to Ref.~\cite{Kopnin1975}. For homogeneous superconductors, the thermal conductivity is smaller than that in the normal state because the quasiparticle has an energy gap $|\Delta|$. At the same time, the condensate cannot carry heat \cite{Geilikman,BardeenReckayzenTewordt}.
The expression for the thermal conductivity is given by \cite{Geilikman}
\begin{equation}
\kappa\propto\int_{|\Delta|}^\infty \frac{{\rm d}\epsilon}{\cosh^2(\epsilon/(2T))}, 
\label{eq: kappa-Geilikman}
\end{equation}
which decreases with increasing $|\Delta|$. Instead of Eq.~\eqref{eq: kappa-Geilikman}, 
we will phenomenologically use a simplified expression for the local thermal conductivity 
 \begin{equation} \label{local_kappa}
   \kappa(x,t) = \kappa_{\textrm{n}} + \left(\kappa_{\textrm{s}} - \kappa_{\textrm{n}} \right) \psi^2(x,t)
 \end{equation}
so that 
$\kappa = \kappa_{\textrm{n}}$ when $\psi = 0$ and $\kappa = \kappa_{\textrm{s}}(<\kappa_{\textrm{n}})$ when $\psi = 1$ respectively, interpolating the two values. Here $\kappa_{\textrm{n}}$ is the thermal conductivity in the normal state, and $\kappa_{\textrm{s}}$ is that in the superconducting state at $T=T_{\rm L}$. 

We can justify the assumption of the local thermal conductivity when the characteristic length scale of spatial variation in the order parameter, namely the coherence length $\xi(T_{\textrm{L}})$, is much larger than the mean free path $l$ or the BCS coherence length $\xi_0 \equiv \hbar v_{\textrm{F}} / (k_{\textrm{B}} T_{\textrm{c}})$, whichever is shorter. This condition generally holds near $T_{\textrm{c}}$. In this parameter regime, the TDGL equation accurately describes the order-parameter dynamics, thereby validating the applicability of our model. Equation~\eqref{local_kappa} is the simplest phenomenological expression that captures the suppression of the thermal conductivity by increasing $|\Delta|$. See the discussion in Sec.~\ref{remarks_tem} for a more general treatment.

The dimensionless expression for Eq.~\eqref{originalTDeq} is given as
\begin{align} \label{TDdimensionless}
\MoveEqLeft{\frac{\partial }{\partial x}\left[ \left\{ 1+(k-1) \psi ^2(x,t) \right\}\frac{\partial \tau(x,t)}{\partial x} \right]}\notag \\
  & = \frac{C} {\kappa_{\textrm{n}} } \frac{\partial \tau(x,t)}{\partial t},
\end{align}
where $k \equiv \kappa_{\textrm{s}} / \kappa_{\textrm{n}}$ is the thermal conductivity ratio for the superconducting and normal states.
We rewrite also the boundary conditions Eqs.~\eqref{bc of T} and~\eqref{bc of OP} as
\begin{subequations} \label{boundconds for gradT}
    \begin{align}
    \tau(x_{\textrm{L}},t) &=\frac{T_{\textrm{L}}}{T_{\textrm{c}}} \equiv \tau_{\textrm{L}} \label{boundconds for gradT first} \\
\tau(x_{\textrm{R}},t) &= \frac{T_{\textrm{R}}}{T_{\textrm{c}}} \equiv \tau_{\textrm{R}} \\
\psi(x_{\textrm{L}},t) &= \tanh{\frac{x_{\textrm{L}}}{\sqrt{2}\xi}} \label{boundconds for gradT third} \\
\psi(x_{\textrm{R}},t) &= \displaystyle \frac{1}{a}\tanh{\frac{x_{\textrm{R}}}{\sqrt{2}a\xi}}. \label{boundconds for gradT last}
    \end{align}
\end{subequations}

We solve the TDGL Eq.~\eqref{TDGLdimensionless} and the thermal diffusion Eq.~\eqref{TDdimensionless} with the boundary conditions Eqs.~\eqref{boundconds for gradT first} -~\eqref{boundconds for gradT last} numerically in Sec. \ref{gradtnum} and analytically in Sec.~\ref{gradtanalitical}.

\subsubsection{Momentum balance relation}  
Equation~\eqref{TDGLdimensionless} yields the following relation:  
\begin{align} 
\MoveEqLeft{\frac{\partial }{\partial x} \left[-\xi^2 \left( \frac{\partial \psi(x,t)}{\partial x} \right)^2 - \frac{\psi^2(x,t)}{1- \tau_{\textrm{L}}}+ \frac{\psi^4(x,t)}{2} \right]}\notag \\
  &+2\tilde{\gamma} \frac{\partial \psi(x,t)}{\partial t}\frac{\partial \psi(x,t)}{\partial x}+ \frac{\tau(x,t)}{1-\tau_{\textrm{L}}}\frac{\partial \psi^2(x,t)}{\partial x}=0,\label{momentum balance relation}
\end{align}
which we can derive by multiplying Eq.~\eqref{TDGLdimensionless} by $\partial \psi/\partial x$ and rewriting so that the resulting equation contains the derivative of the momentum flux tensor.
We call Eq.~\eqref{momentum balance relation} the momentum balance relation because we can derive the same equation also 
from Eq.~(15) in~\cite{KC} or Eq.~(25) in~\cite{Sugai}, which are called `the local balance of the force density', by extracting only the $x$-components, omitting electromagnetic quantities and taking the relaxation coefficient to be real.
The first term on the LHS of Eq.~\eqref{momentum balance relation} represents the divergence of the momentum flux tensor. The second term is the contribution from the relaxation of the order parameter, which gives the dissipative force density. The third term represents the force due to temperature inhomogeneity.

\subsection{numerical calculation} \label{gradtnum}
\begin{widetext}
\begin{figure*}
\includegraphics[width=170mm]{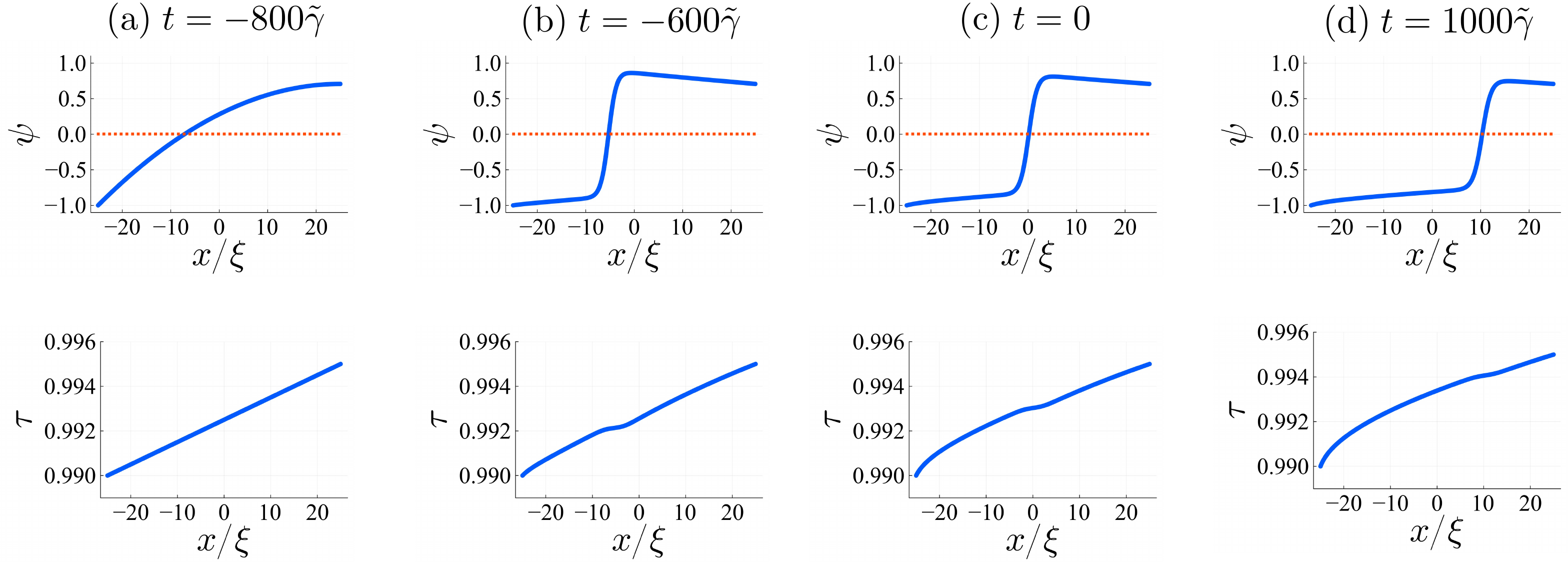}
\captionsetup{justification=raggedright,singlelinecheck=false}
\caption{Snapshots of numerical solutions to the TDGL Eq.~\eqref{TDGLdimensionless} coupled with the thermal diffusion Eq.~\eqref{TDdimensionless}. The figures in the first and second rows show the solutions for the order parameter $\psi(x,t)$ and the temperature $\tau(x,t)$, respectively. The time evolves from (a) to (d). (a): An initial state satisfying the boundary conditions Eqs.~\eqref{boundconds for gradT third} and \eqref{boundconds for gradT last}. 
(b): The initial profile is deformed, and the domain wall-like structure appears. (c), (d): The domain wall keeps flowing against the heat flow, i.e., toward the hotter region.}
\label{figthermalgeadient} 
\end{figure*}
\end{widetext}

We use the fourth-order Runge-Kutta method as a numerical method to update the solution from the parabolic initial state that satisfies the boundary conditions Eqs.~\eqref{boundconds for gradT third} and \eqref{boundconds for gradT last}.  We take the parameters $k =1/20$, $\tilde{\gamma}\kappa_{\textrm{n}}/(C\xi^2)=1$, $\tau_{\textrm{L}} = 0.990$,  $\tau_{\textrm{R}} = 0.995$,  $x_{\textrm{R}}/\xi=25$. Figures~\ref{figthermalgeadient}  (a) - (d) are snapshots arranged in time order. The blue curves in the upper and lower graphs at each time step represent the spatial distributions of the order parameter and the temperature, respectively. The horizontal axis represents the spatial coordinate measured in units of the coherence length. The red dashed lines in the upper panels serve as guides indicating the position of the defect, where the order parameter vanishes. Shortly  ($\sim 5\tilde{\gamma}$) after the initial time,  the order parameter exhibits a characteristic shape around the defect within the range of the coherence length (FIG.~\ref{figthermalgeadient}(b)). This local structure, which we call the domain wall,  keeps moving toward the higher temperature boundary against the heat flow during the motion (FIG.~\ref{figthermalgeadient}(c)(d)). We observe no oscillatory behavior in the domain wall's motion. 
The temperature gradient is locally suppressed in the domain wall region as a consequence of the higher thermal conductivity in the normal region than in the superconducting region.
\subsection{Analytical calculation} \label{gradtanalitical}
The numerical results show that, within our model, the domain wall moves toward the region with the higher temperature. We will discuss this motion based on the analytical solution. In the following section, we consider a flow state in which the domain wall, far from the system boundaries, moves slowly at a constant velocity $v$. The goals of this section are two-fold. (1) To provide explicit definitions of forces on the domain wall. (2) To derive the relation that links the velocity $v$ and the heat flow $q = -\kappa \partial \tau / \partial x$ to explain the direction of motion that the numerical calculation shows.

\subsubsection{Method}
Our strategy follows 
the earlier studies on steady motion of an isolated vortex ~\cite{Schmid66,GK1971,GK1975,Dorsey92}, i.e., 
linearization of the variables with respect to $v$. 
Due care is required in the time domain in the present study, as we consider steady motion in a finite-size system, in contrast to earlier studies. 

The order parameter $\psi(x,t)$ and the temperature $\tau(x,t)$ are expanded as follows with respect to $v$:
\begin{equation} \label{expofpsi}
\psi(x,t)
=\psi_0 (x-vt) + \psi_1(x,t) + \mathcal{O}(v^2)
\end{equation}
and
\begin{align} \label{expoftau}
\begin{split}
\tau(x,t)
&=\tau_0 (x-vt) + \tau_1(x,t) + \mathcal{O}(v^2)
\end{split}\notag
\\&=\tau_{\textrm{L}} + \tau_1(x,t) + \mathcal{O}(v^2).
\end{align}
The subscripts 0 or 1 denote the order of the velocity $v$. Therefore, $\psi_0$ and $\tau_0$ are the equilibrium solutions when we set $\tau_0(x,t) = \tau_{\textrm{L}}$. The explicit expression for $\psi_0(x)$ has been given by Eq.~\eqref{psi0}. We focus on the time domain where 
\begin{equation}
|vt|\ll  \xi \ll x_{\rm R},\label{eq: condition on vt}     
\end{equation}
i.e., the domain wall is far away from the boundaries. 
The solutions to our equations are expressed as the sum of the translation of the solutions in equilibrium: ($\psi_0$, $\tau_0$) and the modulation components: ($\psi_1$, $\tau_1$). The discussion below neglects the terms in $\mathcal{O}(v^2)$. The boundary conditions on $\psi_1$, $\tau_1$ are 
\begin{subequations}    
\begin{align}
\tau_1(x_{\rm R},t)&=\tau_{\rm R}-\tau_{\rm L},\quad \tau_1(x_{\rm L},t)=0,\label{eq: bc-on-tau1}\\
\psi_1(x_{\rm R},t)&=\frac1a\tanh\frac{x_{\rm R}}{\sqrt{2}a\xi}-\tanh\frac{x_{\rm R}-vt}{\sqrt{2}\xi}\notag\\

&=\frac1a\tanh\frac{x_{\rm R}}{\sqrt{2}a\xi}-\tanh\frac{x_{\rm R}}{\sqrt{2}\xi}\nonumber\\
&\quad+\frac{vt}{\sqrt{2}\xi\cosh^2\frac{x_{\rm R}+O(\xi)}{\sqrt{2}\xi}}\label{eq: bc-on-psi1-on-xr-2}\nonumber\\
&\simeq \frac1a-1\\
\psi_1(x_{\rm L},t)&=-\tanh\frac{x_{\rm R}}{\sqrt{2}\xi}+\tanh\frac{x_{\rm R}-vt}{\sqrt{2}\xi}

\nonumber\\
&=-\frac{vt}{\sqrt{2}\xi\cosh^2\frac{x_{\rm R}+O(\xi)}{\sqrt{2}\xi}}\simeq 0
\label{eq: bc-on-psi1-on-xl-2}
\end{align}
\end{subequations}
We thus see that the boundary conditions on $\tau_1$ are independent of $t$ while those on $\psi_1$ can be regarded as $t$-independent under the condition Eq.~\eqref{eq: condition on vt}.   
Substitutions of  Eqs.~\eqref{expofpsi} and~\eqref{expoftau} into Eqs.~\eqref{TDGLdimensionless} and~\eqref{TDdimensionless} yield the relations for each order of the velocity $v$. The zeroth-order relations of these equations are the Ginzburg-Landau equation in equilibrium and a trivial identity. We discuss the first-order relations with respect to $v$ in the following subsections.

\subsubsection{Linearization of the thermal diffusion equation with respect to $v$}
The first-order relation of Eq.~\eqref{TDdimensionless} is given by
\begin{equation} \label{dq/dx=0}
\frac{\partial}{\partial x} \left[\left\{ 1+(k-1)\psi_0^2(x)\right\}\frac{\partial\tau_1(x,t)}{\partial x} \right] = 0.
\end{equation}
In this equation, the operator is independent of $t$. Further, the boundary condition Eq.~\eqref{eq: bc-on-tau1} on $\tau_1$ is $t$-independent. We can thus obtain $t$-independent $\tau_1(x)$ as the solution to Eq.~\eqref{dq/dx=0} under the condition Eq.~\eqref{eq: bc-on-tau1}.  
The integral of Eq.~\eqref{dq/dx=0} yields
\begin{equation} \label{inttau1-with-c}
\tau_1(x)=\mbox{constant}\times \int_{x_{\textrm{L}}}^{x} \frac{\text{d}x'}{1+(k-1)\psi_0^2(x')}.
\end{equation}
 The boundary condition at $x=x_{\rm R}$ in Eq.~\eqref{eq: bc-on-tau1} determines the constant as 
\begin{equation} \label{inttau1-without-c}
\tau_1(x) = (\tau_{\rm R} - \tau_{\rm L}) 
\frac{\displaystyle \int_{x_{\textrm{L}}}^{x} \frac{\text{d}x'}{1+(k-1)\psi_0^2(x')}}
{\displaystyle \int_{x_{\textrm{L}}}^{x_{\rm R}} \frac{\text{d}x'}{1+(k-1)\psi_0^2(x')}} 
\end{equation}
A more explicit expression for $\tau_1(x)$ is available in Appendix \ref{appendix1storder}. 
For later convenience, we rewrite the expression for $\tau_1(x)$ in terms of the heat flow density \eqref{defofq}. By expanding $q(x,t)$ with respect to $v$ in the same way as the case for Eqs.~\eqref{TDGLdimensionless} and \eqref{TDdimensionless}, we obtain
\begin{align}
  \MoveEqLeft{q(x,t) = -\kappa _n T_{\rm c} \left\{1+(k-1)\psi_0^2(x)\right\} \frac{{\rm d}\tau_1(x)}{{\rm d}x}}+ \mathcal{O}(v^2).
  \label{defofq1}
\end{align}
We put the first term on the right-hand side as $q$, which is a constant according to Eq.~\eqref{dq/dx=0}.  From Eqs.~\eqref{inttau1-without-c} and \eqref{defofq1}, we find that

\begin{equation}
  \label{deltatauvsq}
\tau_{\textrm{R}}-\tau_{\textrm{L}} = \frac{-q}{\kappa_{\textrm{n}} T_{\textrm{c}}} \int_{x_{\textrm{L}}}^{x_{\textrm{R}}} \frac{\text{d}x'}{1+(k-1)\psi_0^2(x')},
\end{equation}
which relates $q$ with $\tau_{\textrm{R}}-\tau_{\textrm{L}}$.
With the use of this relation, we obtain the expression for $\tau_1(x)$ in terms of $q$ as

\begin{equation} \label{inttau1}
\tau_1(x)=\frac{-q} {\kappa_{\textrm{n}} T_{\textrm{c}}} \int_{x_{\textrm{L}}}^{x} \frac{\text{d}x'}{1+(k-1)\psi_0^2(x')}.
\end{equation}

\subsubsection{Linearization of the TDGL equation with respect to $v$}\label{subsec: linearTDGL}
The linearization of the TDGL equation (\ref{TDGLdimensionless}) yields 
\begin{equation} \label{tdgl1storder-ux}
  \hat{L} \psi_1(x,t) = \tilde{\gamma}v \frac{\partial \psi_0(u)}{\partial x}- \frac{\psi_0(u)\tau_1(x)}{1- \tau_{\textrm{L}}} , 
\end{equation}
where we introduce the notation $u\equiv x-vt$ and the linear operator $\hat{L} \equiv - \xi^2 \frac{\partial^2}{\partial x^2} -1 + 3\psi_0 ^2(u)$.  Note that
\begin{equation}
  \psi_0(u)=\psi_0(x)(1+O(vt/\xi)).  
\end{equation}
For later purposes, it suffices to obtain $\psi_1(x,t)$  with the accuracy of $O(v)$.  Retaining $O(v)$ terms in Eq.~\eqref{tdgl1storder-ux}, we find that
\begin{equation} \label{tdgl1storder-for-psi1xt}
  \hat{L}_0 \psi_1(x,t)  = \tilde{\gamma}v \frac{\textrm{d} \psi_0(x)}{\textrm{d} x}- \frac{\psi_0(x)\tau_1(x)}{1- \tau_{\textrm{L}}} +O(v^2) 
\end{equation}
with 
\begin{equation}
\hat{L}_0 \equiv - \xi^2 \frac{\textrm{d}^2}{\textrm{d} x^2} -1 + 3\psi_0 ^2(x).
\label{eq: L0}    
\end{equation}
Equation~\eqref{tdgl1storder-for-psi1xt} is a linear differential equation for $\psi_1(x,t)$ with inhomogeneous terms.  The operator $\hat{L}_0$, the inhomogeneous terms in the right-hand side in Eq.~\eqref{tdgl1storder-for-psi1xt}, and the boundary conditions  Eqs.~\eqref{eq: bc-on-psi1-on-xr-2} and \eqref{eq: bc-on-psi1-on-xl-2} are $t$-independent and hence $\psi_1(x,t)$ is also $t$-independent.
The function $\psi_1(x)$  denotes the solution to
\begin{equation} \label{tdgl1storder}
  \hat{L}_0 \psi_1(x)  = \tilde{\gamma}v \frac{\textrm{d} \psi_0(x)}{\textrm{d} x}- \frac{\psi_0(x)\tau_1(x)}{1- \tau_{\textrm{L}}}  
\end{equation}
under the boundary condition Eqs.~\eqref{eq: bc-on-psi1-on-xr-2} and~\eqref{eq: bc-on-psi1-on-xl-2}.  We will show the explicit expression for $\psi_1(x)$  in Appendix~\ref{appendix1storder}. 

By multiplying Eq.~\eqref{tdgl1storder} by $\frac{\textrm{d}\psi_0(x)}{\textrm{d} x}$ and using the equilibrium GL equation $\hat{L}_0 \psi_0(x)=0$,  we obtain 
\begin{align} \label{lmbder0}
  \MoveEqLeft{-\xi^2 \frac{{\rm d}}{{\rm d} x}\left [\frac{{\rm d } \psi_0(x)}{{\rm d } x}\frac{{\rm d }\psi_1(x)}{\partial x}- \frac{{\rm d }^2\psi_0(x)}{{\rm d } x^2} \psi_1(x)\right]} \notag \\& -\tilde{\gamma}v \left( \frac{{\rm d }\psi_0(x)}{{\rm d } x} \right) ^2 + \frac{\tau_1(x)}{2(1- \tau_{\textrm{L}})}\frac{{\rm d }}{{\rm d } x}\psi_0^2(x) =0.
\end{align}
Equation~\eqref{lmbder0} is identical to the linearization of Eq.~\eqref{momentum balance relation}, and thus we regard Eq.~\eqref{lmbder0} as the linearized local momentum balance relation. 

\subsubsection{Forces on the domain wall}
As mentioned in Subsection \ref{subsec: linearTDGL}, we can regard Eq.~\eqref{lmbder0} as the local momentum balance relation, and thus, each term on the LHS represents force density. In this subsection, we identify the integral of these terms as the driving force, the viscous force, and the thermal force, respectively. 
\paragraph{Driving force}
The first term on the LHS of Eq.~\eqref{lmbder0} is a total derivative and its integral over the interval gives the net momentum that enters and exits through the boundaries, namely the driving force\cite{Sugai}. We define the driving force as
\begin{align}
\label{Fdrv}
\begin{split}
F_{\textrm{drv}}
& \equiv
- \xi^2\int_{x_{\textrm{L}}}^{x_{\textrm{R}}}\text{d}x \    \frac{{\rm d}}{{\rm d} x}\left [ \frac{\text{d}\psi_0(x)}{\text{d}x} \frac{{\rm d} \psi_1(x)}{\partial x}  \notag \right. \\& \left. \hspace{1.9cm} - \frac{\text{d}^2\psi_0(x)}{{\rm d} x^2}\psi_1(x)\right]
\end{split} \notag
\\&=
-\xi^2 \left[ \frac{\text{d}\psi_0(x)}{\text{d}x} \frac{{\rm d}\psi_1(x)}{{\rm d} x}  - \frac{\text{d}^2\psi_0(x)}{{\rm d} x^2}\psi_1(x) \right]_{x_{\textrm{L}}}^{x_{\textrm{R}}}.
\end{align}

\paragraph{Viscous force} 
The second term on the LHS of Eq.~\eqref{lmbder0} has the sign opposite to $v$, so that we can regard the integral as a viscous force on the domain wall 
\begin{equation}
\label{Fvis}
F_{\textrm{vis}} \equiv -\tilde{\gamma}v\int_{x_{\textrm{L}}}^{x_{\textrm{R}}} \text{d}x \    \left( \frac{\text{d}\psi_0(x)}{\text{d}x} \right)^2.
\end{equation}
\paragraph{Thermal force}
We define the thermal force as the integral of the third term on the LHS of Eq.~\eqref{lmbder0} as
\begin{equation} \label{Fth}
F_{\textrm{th}} \equiv  \frac{1}{2(1- \tau_{\textrm{L}})}\int_{x_{\textrm{L}}}^{x_{\textrm{R}}} \text{d}x \tau_1(x)\frac{\text{d}}{\text{d}x}\psi_0^2(x),
\end{equation}
which we can rewrite as
\begin{subequations}    
\begin{align}
  F_{\textrm{th}}&=\frac{1}{2(1- \tau_{\textrm{L}})}\int_{x_{\textrm{L}}}^{x_{\textrm{R}}} \text{d}x    \tau_1(x)\frac{\text{d}}{\text{d}x}(\psi_0^2(x)-\psi_0^2(x_{\rm R}))\label{eq: Fth-1}\\
  &= \frac{1}{2(1-\tau_{\textrm{L}})}\int_{x_{\textrm{L}}}^{x_{\textrm{R}}} \text{d}x    \frac{\text{d}\tau_1(x)}{\text{d}x}(\psi_0^2(x_{\rm R})-\psi_0^2(x))\label{eq: Fth-2}\\
  &= \frac{-q}{2\left( 1-\tau_{\textrm{L}}\right)\kappa_{\textrm{n}}T_{\textrm{c}}}\int_{x_{\textrm{L}}}^{x_{\textrm{R}}} \text{d}x \    \frac{\psi_0^2\left( x_{\textrm{R}}\right)-\psi_0^2\left( x\right)}{1+(k-1)\psi_0^2\left( x\right)}\label{Fthq}.
\end{align}
\end{subequations}
We perform the partial integral from Eq.~\eqref{eq: Fth-1} to Eq.~\eqref{eq: Fth-2}. We use Eq.~\eqref{inttau1} from Eq.~\eqref{eq: Fth-2} to Eq.~\eqref{Fthq}.

When $x_{\rm R}/\xi=|x_{\rm L}|/\xi\gg 1$,   $F_{\textrm{drv}}$ vanishes, and $F_{\textrm{vis}}$ and  $F_{\textrm{th}}$ are balanced, i.e.,
\begin{equation}
\label{forcebalance}
F_{\textrm{vis}}+F_{\textrm{th}}= 0, 
\end{equation}
In the following discussion, we neglect the driving force and assume Eq.~\eqref{forcebalance} holds. However, we note that the driving force can be crucial in a vortex lattice or in a finite system where the system size is comparable to the coherence length.
\subsubsection{Linear relation between $v$ and $q$}
The force-balance relation Eq.~\eqref{forcebalance} 
 yields the linear relation between $v$ and $q$ as
\begin{align}
v &= \frac{\displaystyle \frac{1}{2(1-\tau_{\textrm{L}})}\int_{x_{\textrm{L}}}^{x_{\textrm{R}}} \text{d}x    \frac{\text{d}\tau_1(x)}{\text{d}x}(\psi_0^2(x_{\rm R})-\psi_0^2(x))}{\displaystyle \int_{x_{\textrm{L}}}^{x_{\textrm{R}}} \text{d}x \ \left( \frac{\text{d}\psi_0(x)}{\text{d}x} \right) ^2} \nonumber \\[\jot]
 & = \frac{-q G[\psi_0]}{2(1-\tau_{\textrm{L}}) \tilde{\gamma} \kappa_{\textrm{n}} T_{\textrm{c}}} = \frac{-qG[\psi_0]}{2\gamma \kappa_{\textrm{n}} T_{\textrm{c}}} , \label{vvsqtransport}
\end{align}
where
\begin{equation} \label{defofG}
   G[\psi_0] \equiv \frac{\displaystyle \int _{x_{\textrm{L}}}^{x_{\textrm{R}}}\text{d}x \ \frac{\psi_0^2\left( x_{\textrm{R}}\right)-\psi_0^2\left( x\right)}{1+(k-1)\psi_0^2(x)}}{\displaystyle \int_{x_{\textrm{L}}}^{x_{\textrm{R}}} \text{d}x \ \left( \frac{\text{d}\psi_0(x)}{\text{d}x} \right) ^2}.
\end{equation}
The functional $G$ is always positive when $k=\kappa_{\textrm{s}}/\kappa_{\textrm{n}}$ is smaller than $1$ because $\psi_0^2 (x) \leq \psi_0^2 (x_{\textrm{R}}) \leq 1$. Hence, Equation~\eqref{vvsqtransport} shows that $v$ and $q$ have different signs. In the limit of $x_{\rm R}/\xi\rightarrow \infty$ and $x_{\rm L}/\xi\rightarrow -\infty$, we have 
\begin{equation}
 G[\psi_0]\rightarrow \frac{6\xi^2}{|1-k|^\frac12}\ln\left(\frac{1+|1-k|^\frac12}{1-|1-k|^\frac12}\right)
\end{equation}
and
\begin{equation}
 v\rightarrow \frac{-3q\xi^2}{\gamma \kappa_{\textrm{n}} T_{\textrm{c}}|1-k|^\frac12}\ln\left(\frac{1+|1-k|^\frac12}{1-|1-k|^\frac12}\right). 
\end{equation}

In conclusion, the domain wall moves from the lower- to the higher-temperature region, not along the heat flow but against it. This result is consistent with the results obtained from the numerical calculation shown in Sec.~\ref{gradtnum}. We can thus regard Eq.~\eqref{vvsqtransport} as the theoretical basis for this phenomenon under conditions where the temperature gradient or heat flow is sufficiently small. 

\subsubsection{Comparison with numerical results}\label{comp.gradT}
\begin{figure}
\includegraphics[pagebox=artbox,width=60mm]{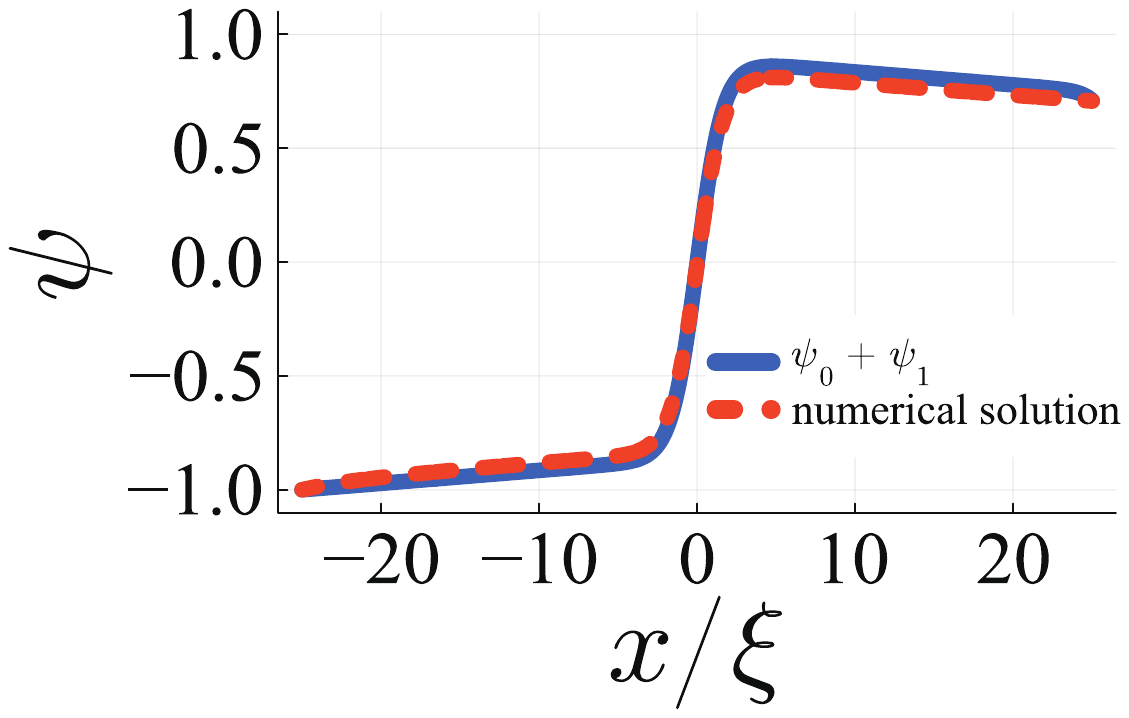}
\captionsetup{justification=raggedright,singlelinecheck=false}
\caption{Comparison between the analytic solution up to the first order, namely $\psi_0(x) + \psi_1(x)$ plotted in the blue solid curve, and the snapshot at the time $t=0$ in the time-evolution obtained from the numerical calculation we discuss in Sec.~\ref{gradtnum} plotted in the red dashed curve.} 
\label{fig1st-psi0+psi1} 
\end{figure}

We show the consistency between the numerical and the analytical results. Figure~\ref{fig1st-psi0+psi1} shows that the function $\psi_0(x) + \psi_1(x)$ is a good approximation to the numerical result $\psi(x,t)$ at the time  $t=0$,  when the domain wall passes through $x=0$.  
Profiles of  the numerical solutions to Eqs.~\eqref{dq/dx=0} and~\eqref{tdgl1storder}, i.e., $\tau_1$ and $\psi_1$ are available in \ref{comp.gradT} in Appendix~\ref{appendix1storder}. 

Figure~\ref{fig: x-t-curve} compares the time dependence of the domain wall coordinate obtained from the numerical calculation with the result of the linear analysis (Eq.~\eqref{vvsqtransport}) under the same parameters. Here, the high-temperature boundary condition is $\tau_{\rm R} = 0.990+10^{-6}$, and the other parameters are equal to the values we adopted in FIG.~\ref{figthermalgeadient}. We used a different value of $\tau_{\rm R}$ to validate the linearity of $\psi(x,t)$ and $\tau(x,t)$ with respect to the velocity of the domain wall. If the temperature difference at the boundaries is small enough, the velocity of the domain wall obtained from numerical calculations agrees well with the result of the linear analysis. As the temperature difference increases, the non-linearity becomes non-negligible, and the ratio between them deviates from 1. For the dependence of this ratio on temperature difference, see Supplemental Material~\cite{sm_ratio_vana_vnum}.

\begin{figure}
\includegraphics[width=70mm]{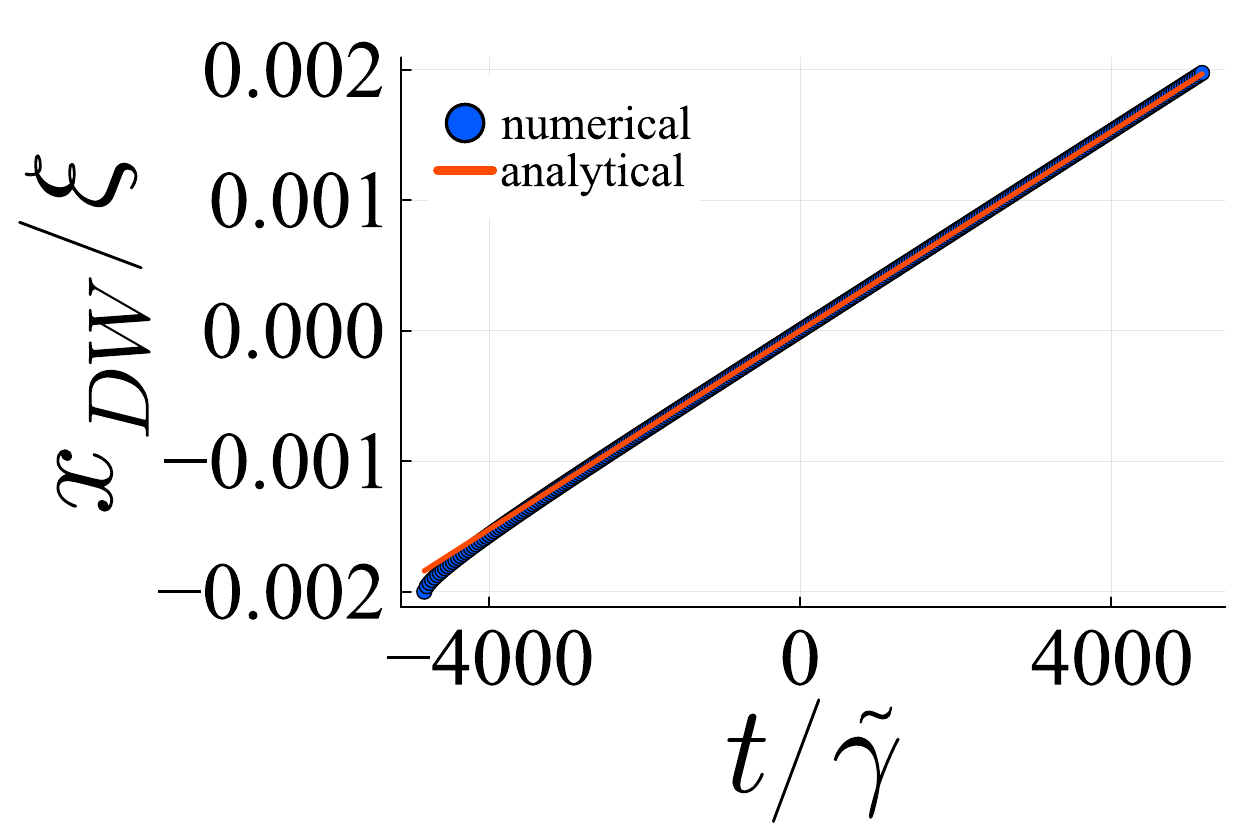}
\captionsetup{justification=raggedright,singlelinecheck=false}
\caption{Time-dependence of the coordinate of the domain wall. The solid red line and the blue curve represent the analytical and numerical result under the parameters $k =1/20$, $\tilde{\gamma}\kappa_{\textrm{n}}/(C\xi^2)=1$, $\tau_{\textrm{L}} = 0.990$,  $\tau_{\textrm{R}} = 0.990+10^{-6}$,  $x_{\textrm{R}}/\xi=25$, respectively. The ratio between the analytically obtained velocity $v_{\rm ana}$ and the numerically obtained velocity $v_{\rm sim}$ is $v_{\rm sim} / v_{\rm ana}=1.0039$ (see Supplemental Material\cite{sm_ratio_vana_vnum} for temperature dependency of the ratio.). }
\label{fig: x-t-curve}
\end{figure}

\subsection{Remarks} \label{remarks_tem}
\begin{figure*}
\includegraphics[width=110mm]{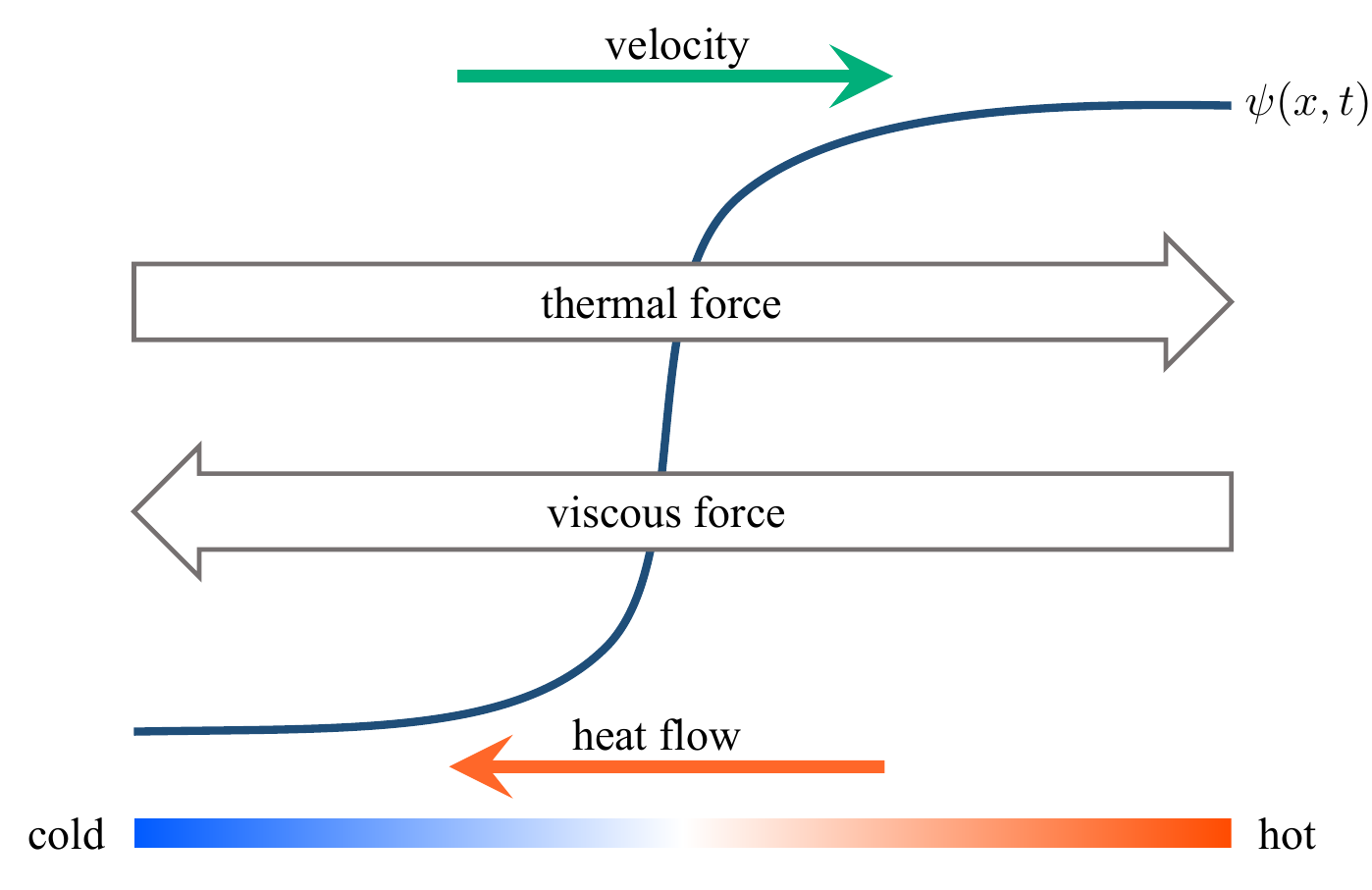}
\captionsetup{justification=raggedright,singlelinecheck=false}
\caption{Illustration of the relations among the heat flow, velocity, and forces. According to Eq. (\ref{vvsqtransport}), the velocity of the domain wall is directed to the hotter region since the heat flows toward the colder region. The direction of the thermal force follows from  Eq.~\eqref{Fthq}.  The viscous force, which is against the velocity, is balanced with the thermal force. The driving force is negligible when the system size is much larger than $\xi$, i.e., $|x_{\rm L}|,x_{\rm R}\gg \xi$.}
\label{figqvforces} 
\end{figure*}

We easily see the signs of the domain wall's velocity $v$, the heat flow $q$, and forces $F_{\textrm{vis}}$, $F_{\textrm{th}}$. Equation~\eqref{deltatauvsq} tells us $q<0$, so $v>0$ since they have different signs. Obviously, $F_{\textrm{vis}}<0$ as it is against $v$. Equation~\eqref{Fthq} shows $F_{\textrm{th}}>0$. We summarize these results in FIG. \ref{figqvforces}.

We can understand the domain wall's motion toward the higher temperature in the same way as the transient process of vortex pinning, where defects or impurities locally suppress the order parameter.

\begin{figure*}
\includegraphics[width=110mm]{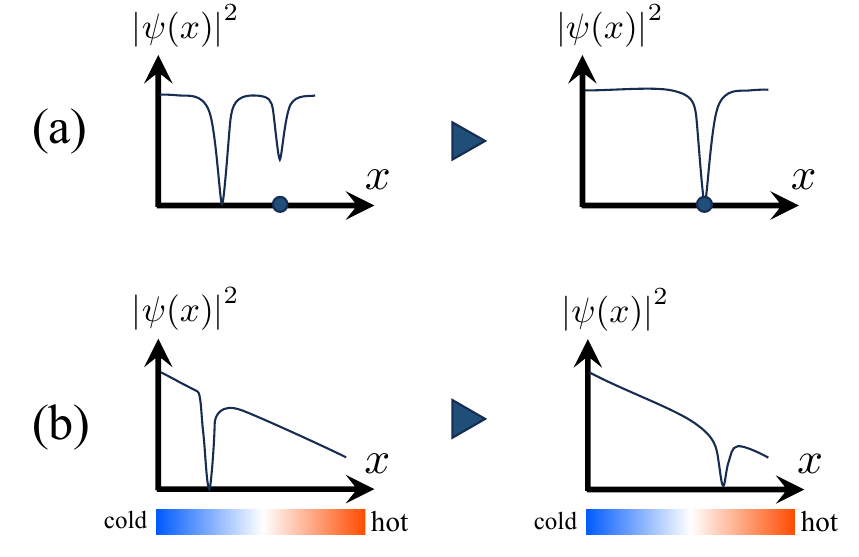}
\captionsetup{justification=raggedright,singlelinecheck=false}
\caption{A comparison between the motion of the order parameters. (a) In the pinning effect, defects tend to move toward pinning centers, such as impurities, represented by the dot on the horizontal axis. (b) The motion of the domain wall under a temperature gradient has been discussed so far. The similarity between the two cases lies in the same energetic tendency: the defect moves toward a region where the order parameter is suppressed, thereby reducing the loss of condensation energy.}
\label{figpinninggradT} 
\end{figure*}

Vortices are
energetically favored in regions where the order parameter is suppressed. Therefore, such defects and impurities attract vortices. A similar
energetic argument applies in the presence of a temperature gradient:
the domain wall is attracted toward the hotter region, where the
amplitude of the order parameter is smaller than in the colder region
(Fig.~\ref{figpinninggradT}).
\color{black}

At the end of this section, we refer to more general possibilities of Eq.~\eqref{local_kappa}. Consider the case when we assume $\kappa$ is a monotonically decreasing functional of $\left| \psi \right|$, namely, $\kappa[\psi]$ instead of Eq.~\eqref{local_kappa}. Then the corresponding expression for the velocity of the domain wall is given by
\begin{equation} \label{vG'}
  v = \frac{-qG'[\psi_0]}{2\gamma \kappa_{\textrm{n}} T_{\textrm{c}}},
\end{equation}
where
\begin{equation}
  G'[\psi_0] \equiv \frac{\displaystyle \int _{x_{\textrm{L}}}^{x_{\textrm{R}}}\text{d}x \ \frac{\psi_0^2\left( x_{\textrm{R}}\right)-\psi_0^2\left( x\right)}{\kappa[\psi_0]/\kappa_{\textrm{n}}}}{\displaystyle \int_{x_{\textrm{L}}}^{x_{\textrm{R}}} \text{d}x \ \left( \frac{\text{d}\psi_0(x)}{\text{d}x} \right) ^2}.
\end{equation}
The spatial distribution of temperature may differ from that obtained under the previously discussed model. However, Eq.~\eqref{vG'} shows the sign of the velocity remains positive, indicating that the motion still occurs toward the higher temperature region. Therefore, although the assumption expressed in Eq.~\eqref{local_kappa} is not strictly necessary, we adopt it in this paper to facilitate the clarity of the discussion and to derive the explicit expression. This general conclusion holds even for a uniform thermal conductivity ($\kappa(x,t) = \textrm{const.} $, namely $k \rightarrow 1$), where the sign of the velocity remains positive.

\section{Motion of domain wall under spin-density gradient} \label{sec3gradmu}
In this section, we address the motion of domain wall under a spin-accumulation gradient. As in the previous section, we consider a superconductor with spatially varying physical quantities only in the $x$-coordinate in the finite region $x_{\textrm{L}} \leq x\leq x_{\textrm{R}}$ and assume that the system is uniform in the $yz$ plane.  We set the linear dimension $x_{\rm R}-x_{\rm  L}$  to be much larger than $\xi$ but smaller or comparable to the spin diffusion length.  

The model comprises the TDGL equation and the spin-diffusion equation. In the TDGL equation, we take into account the local dependence of the critical temperature on the modulus of the spin accumulation~\cite{Takahashi1999,Vargas2020}.  In the spin diffusion equation, we consider the local dependences of the spin conductivity and spin relaxation time on the order parameter.  Through these mutual dependences between the spin and the superconducting properties, the TDGL and spin diffusion equations are coupled.  The microscopic derivation of the spin diffusion equation in superconductors is available in~\cite{Taira2018}. 

\subsection{Model}
We address the motion of the domain wall in the presence of  the spin accumulation $\mu(x,t)$ by including the $\mu(x,t)$-dependence of the critical temperature $T_{\textrm{c}}\left(\mu^2(x,t)\right)$ in the TDGL equation
\begin{align} \label{eq: TDGLspin-dimensionful}
-\frac{\hbar^2}{2m}\MoveEqLeft{\frac{\partial^2 \Psi(x,t)}{\partial x^2} +\alpha_0\left(\frac{T}{T_{\textrm{c}}( \mu^2(x,t))}-1\right)\Psi(x,t)}\notag\\
  &+ \beta \Psi^3 (x,t)
  =-\tilde{\gamma} \frac{\partial \Psi(x,t)}{\partial t}.
\end{align}

Introducing the dimensionless order parameter $\psi$, the coherence length $\xi$, and the normalized  relaxation coefficient $\tilde{\gamma}$ as
\begin{equation}
\psi(x,t) \equiv \frac{\displaystyle \Psi(x,t)}{\sqrt{\frac{\displaystyle \alpha_0 \left( T_{\textrm{c}}\left(0\right)-T \right)}{\displaystyle\beta T_{\textrm{c}}\left(0\right)}}}
\end{equation}
\begin{equation}
\xi \equiv \sqrt{\frac{\hbar^2 T_{\textrm{c}}\left(0\right)}{2m \alpha_0 \left( T_{\textrm{c}}\left(0\right)-T\right)}}
\end{equation}
\begin{equation}
\tilde{\gamma} \equiv  \frac{\gamma}{\alpha_0}\frac{T_{\textrm{c}}\left( 0\right)}{T_{\textrm{c}}\left( 0\right)-T},
\end{equation}
we rewrite the TDGL in terms of $\psi(x,t)$ and $\mu^2(x,t)$,
\begin{align} \label{TDGLspin}
\MoveEqLeft{-\xi^2 \frac{\partial^2 \psi}{\partial x^2} +\left(\frac{T}{T_{\textrm{c}}\left( \mu^2\right)}-1\right)\left(1-\frac{T}{T_{\textrm{c}}\left( 0\right)}\right)^{-1}\psi}+ \psi^3\notag \\
  & =-\tilde{\gamma} \frac{\partial \psi}{\partial t}.
\end{align}
Note that the temperature $T$ is constant, in contrast to the model discussed in the previous section.

In addition, we adopt the spin diffusion equation
\begin{equation}
\label{SDeq}
\frac{\partial }{\partial x}\left[\sigma^{\textrm{sp}}(x,t)\frac{\partial \mu(x,t)}{\partial x} \right]-\frac{\mu(x,t)}{\tau^{\rm sp}(x,t)}=\frac{\partial \mu(x,t)}{\partial t}.
\end{equation}
Here,  $\sigma^{\textrm{sp}}(x,t)$, $\tau^{\rm sp}(x,t)$ represent the spin conductivity and the spin relaxation time. Again, as we did before, we assume that these parameters spatially and temporally vary via the order parameter dependency given by
\begin{subequations}
    \begin{align}
    \displaystyle
    \sigma^{\textrm{sp}}(x,t) = \sigma^{\textrm{sp}}_{\textrm{n}}+\left( \sigma^{\textrm{sp}}_{\textrm{s}}-\sigma^{\textrm{sp}}_{\textrm{n}}\right)\psi^2(x,t)  \\
    \displaystyle
   \tau^{\rm sp}(x,t) = \tau^{\rm sp}_{\textrm{n}}+\left( \tau^{\rm sp}_{\textrm{s}}-\tau^{\rm sp}_{\textrm{n}}\right)\psi^2(x,t)
    \end{align}
\end{subequations}
so that the values for the superconducting (with subscript $\text{s}$) and normal (with subscript $\text{n}$) states are interpolated.

Taking $\mu^2(x,t)$, instead of $\mu(x,t)$, as a new variable, we can rewrite the spin diffusion equation Eq.~\eqref{SDeq} into the following form:
\begin{align} \label{spindiffusionmu2}
\MoveEqLeft{\frac{\partial }{\partial x}\left[\sigma^{\textrm{sp}}(x,t)\frac{\partial \mu^2(x,t)}{\partial x} \right]-\frac{\sigma^{\textrm{sp}}(x,t)}{2\mu^2(x,t)}\left(\frac{\partial \mu^2(x,t)}{\partial x}\right)^2}\notag \\
  &-\frac{2\mu^2(x,t)}{\tau^{\rm sp}(x,t)}=\frac{\partial \mu^2(x,t)}{\partial t}.
\end{align}
We impose the boundary conditions for the spin accumulation as

\label{bcformu}
    \begin{align}
    \displaystyle
    \mu^2(x_{\textrm{L}},t) &= 0,\quad 
    \mu^2(x_{\textrm{R}},t) = \mu^2_{\textrm{R}}

  \label{eq: bcformu} 
    \end{align}

and those for the dimensionless order parameter are given as
\begin{subequations} \label{bcforpsiforspin}
    \begin{align}
    \displaystyle
    \psi(x_{\textrm{L}},t) &= \tanh{\frac{x_{\textrm{L}}}{\sqrt{2}\xi}} \label{bcforpsiforspinleft}, \\
    \displaystyle
   \psi(x_{\textrm{R}},t) &= \displaystyle \frac{1}{a^{\prime}}\tanh{\frac{x_{\textrm{R}}}{\sqrt{2}a^{\prime}\xi}}
, \label{bcforpsiforspinright}
    \end{align}
\end{subequations}
where we introduce the constant 
\begin{equation}
a^{\prime} \equiv  \displaystyle \sqrt{\frac{T_{\textrm{c}}\left(\mu^2_{\textrm{R}}\right)}{T_{\textrm{c}}\left(\mu^2_{\textrm{R}}\right)-T}\frac{T_{\textrm{c}}\left( 0\right)-T}{T_{\textrm{c}}\left( 0\right)}}.
\label{eq: aprime}
\end{equation}

The momentum balance relation in the present setup is

\begin{align} 
\MoveEqLeft{\frac{\partial }{\partial x} \left[-\xi^2 \left( \frac{\partial \psi(x,t)}{\partial x} \right)^2 + \frac{\psi^2(x,t)}{a'^2}+ \frac{\psi^4(x,t)}{2} \right]}\notag \\
  &+2\tilde{\gamma} \frac{\partial \psi(x,t)}{\partial t}\frac{\partial \psi(x,t)}{\partial x}+b\psi^2(x,t) \frac{\partial \mu^2(x,t)}{\partial x}=0,\label{momentum balance relation-in-spin}
\end{align}
which we can use to identify the force due to the spin-accumulation gradient. Here we define the symbol $b$ as
\begin{equation}
b \equiv \frac{T}{T_{\textrm{c}}\left( 0\right)\left(T_{\textrm{c}}\left( 0\right)-T\right)}\left. \frac{\text{d}T_{\textrm{c}}\left( \mu^2\right)}{\text{d}\mu^2}\right|_{\mu^2=0} < 0.
\end{equation}

To sum up, we numerically and analytically solve the TDGL equation Eq.~\eqref{TDGLspin} and the spin diffusion equation  Eq.~\eqref{spindiffusionmu2}  in terms of $\mu^2(x,t)$ under the boundary conditions Eqs.~\eqref{eq: bcformu}, ~\eqref{bcforpsiforspinleft}, and~\eqref{bcforpsiforspinright}.

\subsection{Numerical calculation}\label{grad_spin_num}
We solve Eqs.~\eqref{TDGLspin} and~\eqref{spindiffusionmu2} under the boundary condition Eqs.~\eqref{eq: bcformu} and~\eqref{bcforpsiforspin} numerically by the fourth-order Runge-Kutta method. We take the parameters as, $x_{\textrm{R}}/\xi=25$, $\sigma_{\textrm{s}}/\sigma_{\textrm{n}}=\tau_{\textrm{s}}/\tau_{\textrm{n}}=2$, $\xi^2/(\sigma_{\textrm{n}}\tau_{\textrm{n}})=1/1000$, $\xi^2/(\sigma_{\textrm{n}}\tilde{\gamma})=10$, $T=0.99T_{\textrm{c}}(\mu^2_{\textrm{R}})$ and $\mu_{\rm R}/\Delta_0 = 0.20$, where $\Delta_0$ is the absolute value of the superconducting gap at absolute zero. 
We found the transition temperature $T_{\textrm{c}}\left(\mu^2\right)$ by numerically solving the gap equation, which takes into account the difference between the chemical potentials of electrons with different spins ~\cite{Takahashi1999,Yang2010}. 

\begin{figure*}
\includegraphics[width=170mm]{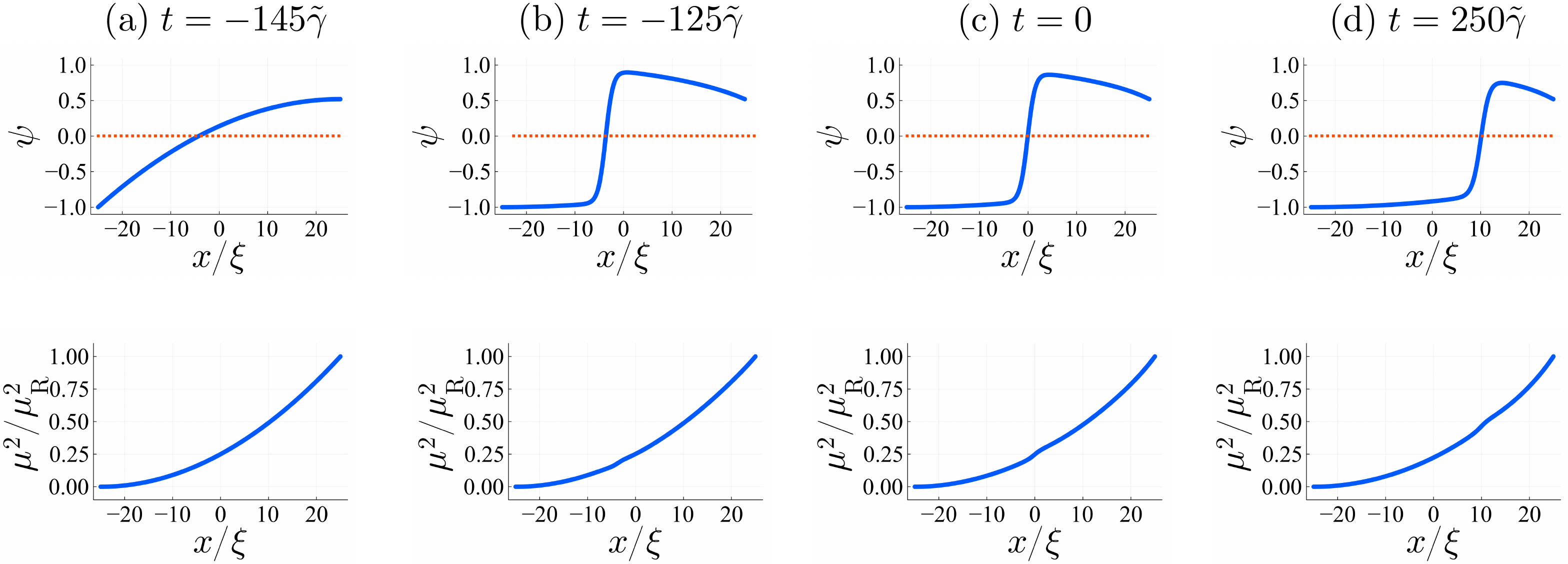}
\captionsetup{justification=raggedright,singlelinecheck=false}
\caption{
Snapshots of numerical solutions to the TDGL Eq.~\eqref{TDGLdimensionless} 
coupled with the spin diffusion equation Eq.~\eqref{spindiffusionmu2}. The figures in the first and second rows show the solutions for the order parameter $\psi(x,t)$ and the modulus of spin accumulation $\mu^2(x,t)$, respectively.
The time evolves from (a) to (d). (a) An 
initial state satisfying the boundary conditions.
(b) Immediately after the initial time, the initial profile is deformed so that the domain wall-like structure appears. (c), (d) The domain wall moves to the region where the modulus of the spin accumulation is larger.}
\label{figspingradient} 
\end{figure*}

Figure~\ref{figspingradient} shows numerical results, where  (a), (b), (c), and (d) are arranged in time order. 
The upper and lower panels for each time step show the spatial distribution of the order parameter and the squared modulus of the spin accumulation divided by $\mu_{\textrm{R}}^2$, respectively. The horizontal and vertical axes and the red dashed line represent the same as those in FIG. \ref{figthermalgeadient}.

As in the case of the temperature gradient shown in FIG. \ref{figthermalgeadient}, we found that the domain wall moves toward the boundary where the modulus of the spin accumulation is larger.
\subsection{Analytical calculation}
We consider the domain wall around $x=0$ moving at a velocity $v$ in the time domain satisfying  $|vt|/\xi\ll 1$. 
\subsubsection{Linearization with respect to $v$}
Following the same procedure explained in Sec.~\ref{gradtanalitical}, we linearized the spin diffusion and the TDGL equation with respect to $v$ as,
\begin{subequations}
\label{eq: LRT-psi-mu}    
\begin{align} \label{expofmu}
\psi(x,t)&=\psi_0(x-vt)+\psi_1(x)+\mathcal{O}(v^2)\\   

\mu^2(x,t)
&=\mu^2_0 (x-vt) + \mu^2_1(x) + \mathcal{O}(v^2)\notag\\

&= \mu^2_1(x) + \mathcal{O}(v^2).\label{eq: LRT-mu-2}
\end{align}
\end{subequations}

No term in the $0$-th order appears in Eq.~\eqref{eq: LRT-mu-2} since there is no spin accumulation in equilibrium. $\psi_1(x)$ and $\mu^2_1(x)$ satisfy the boundary conditions 
\begin{subequations} \label{bcformu-1}
    \begin{align}

    \psi_1(x_{\textrm{L}}) &= 0,\quad   \psi_1(x_{\textrm{R}}) =\frac{1}{a'}-1\label{bcformuleft-1}\\
\displaystyle
    \mu_1^2(x_{\textrm{L}}) &= 0,\quad   \mu_1^2(x_{\textrm{R}}) = \mu^2_{\textrm{R}}
\label{bcformuright-1}
    \end{align}
\end{subequations}
with $a'$ defined in Eq.~\eqref{eq: aprime}.  The linearized equations are 

\begin{align}
\label{linearlizedSDeq}
\MoveEqLeft{\frac{\text{d}}{\text{d} x}\left[\sigma^{\textrm{sp}}_0(x)\frac{\text{d}\mu^2_1(x)}{\text{d}x} \right]-\frac{\sigma^{\textrm{sp}}_0(x)}{2\mu^2_1(x)}\left(\frac{\text{d}\mu^2_1(x)}{\text{d}x}\right)^2}\notag \\
  &-\frac{2\mu^2_1(x)}{\tau^{\rm sp}_0(x)}=0
\end{align}

and
\begin{equation}
\hat{L}_0\psi_1(x)=b\mu_1^2(x)\psi_0(x)+\tilde{\gamma}v\frac{{\rm d}\psi_0(x)}{{\rm d}x},    \label{eq: linearization-of-L0x}
\end{equation}
with  $\hat{L}_0$ being defined in Eq.~\eqref{eq: L0}.
Here we introduce the following notations:\begin{subequations}
    \begin{align}
    \displaystyle
    \sigma^{\textrm{sp}}_0(x)& = \sigma^{\textrm{sp}}_{\textrm{n}}+\left( \sigma^{\textrm{sp}}_{\textrm{s}}-\sigma^{\textrm{sp}}_{\textrm{n}}\right)\psi_0^2(x)  \\
    \displaystyle
   \tau^{\rm sp}_0(x)& = \tau^{\rm sp}_{\textrm{n}}+\left( \tau^{\rm sp}_{\textrm{s}}-\tau^{\rm sp}_{\textrm{n}}\right)\psi_0^2(x).
    \end{align}
\end{subequations}

We can derive 
\begin{align} \label{linearlizedtdglspin}
\MoveEqLeft{- \xi^2 \frac{\text{d}}{\text{d} x}\left[\frac{\text{d}\psi_0(x)}{\text{d}x} \frac{\text{d} \psi_1(x)}{\text{d} x}- \frac{\text{d}^2\psi_0(x)}{\text{d}x^2}\psi_1(x)\right]}\notag \\
  &-\tilde{\gamma}v \left(\frac{\text{d}\psi_0(x)}{\text{d}x}\right)^2 -\frac{b}{2}\mu^2_1(x) \frac{\text{d}}{\text{d}x}\psi^2_0(x) =0
\end{align}
from the linearization of Eq.~\eqref{momentum balance relation-in-spin}  or from Eq.~\eqref{eq: linearization-of-L0x} with the use of  Eq.~\eqref{eqgl0}. We can read Eq.~\eqref{linearlizedtdglspin} as the local momentum balance relation. The first and second terms are identical to the ones in Eq.~\eqref{lmbder0}. The third term represents a force arising from the spin-density gradient.

\subsubsection{Forces on domain wall}
We can classify forces on the domain wall into three types based on their origins. The definitions of $F_{\textrm{drv}}$ and $F_{\textrm{vis}}$ require no corrections to Eqs.~\eqref{Fdrv} and \eqref{Fvis} respectively.

We define the force due to the spin accumulation gradient as
\begin{align}

\label{Fenvspin}
F_{\textrm{spin}}
& \equiv -\frac{b}{2} \int_{x_{\textrm{L}}}^{x_{\textrm{R}}}\text{d}x \ \mu^2_1(x) \frac{\text{d}}{\text{d}x}\psi^2_0(x) \notag,

\end{align}
which we can rewrite as
\begin{equation}
F_{\textrm{spin}}=
-\frac{b}{2}\int_{x_{\textrm{L}}}^{x_{\textrm{R}}}\text{d}x \ \left(\psi^2_0(x_{\textrm{R}})-\psi^2_0(x)\right)\frac{\text{d}\mu^2_1(x)}{\text{d}x} 
\label{Fenvspin}
\end{equation}
in a way similar to Eq.~\eqref{eq: Fth-2}. 
The factor $\psi^2_0(x_{\textrm{R}})-\psi^2_0(x)$ in the integrand of Eq.~\eqref{Fenvspin} makes the force density concentrated around the domain wall. We can thus regard Eq.~\eqref{Fenvspin} as a force acting on the domain wall. We find that $F_{\textrm{spin}}>0$ because \begin{equation} \label{mu2mon}
0 \leq \frac{\text{d}\mu^2_1(x)}{\text{d}x},
\end{equation}

as shown in Appendix \ref{appendixmu12deriv}. 
We are now ready to determine the sign of the velocity, namely, the direction of the domain wall's motion. According to Eq.~\eqref{linearlizedtdglspin}, we have the force-balance relation $F_{\textrm{drv}}+F_{\textrm{vis}}+F_{\textrm{spin}}=0$ for the finite system. As in the previous setup, the driving force vanishes in the limits $x_{\rm R}/\xi\rightarrow \infty$ and $x_{\rm L}/\xi\rightarrow -\infty$. Therefore, we obtain
\begin{equation}
F_{\textrm{vis}}+F_{\textrm{spin}}=0.
\end{equation}
This force-balance relation and the fact that $F_{\textrm{spin}}>0$ directly yield $F_{\textrm{vis}}<0$. Because of Eq.~\eqref{Fvis}, the signs of the viscous force and the velocity are the opposite, $v>0$. The explicit expression for $v$ is 
\begin{equation}\label{vvsdmu2dx}
  v = 
\frac{
  \displaystyle
  -\frac{b}{2} \int_{x_{\textrm{L}}}^{x_{\textrm{R}}} \text{d}x \;
  \frac{\text{d} \mu^2_1(x)}{\text{d}x}
  \left( \psi^2_0(x_{\textrm{R}}) - \psi^2_0(x) \right)
}{
  \displaystyle
   \tilde{\gamma} \int_{x_{\textrm{L}}}^{x_{\textrm{R}}} \text{d}x \;
  \left( \frac{\text{d} \psi_0(x)}{\text{d}x} \right)^2
} > 0
\end{equation}
i.e., the domain wall moves toward the region where the modulus of the spin accumulation is larger.    

\subsubsection{Comparison with numerical results}\label{compgradmu}
\begin{figure}
\includegraphics[pagebox=artbox,width=60mm]{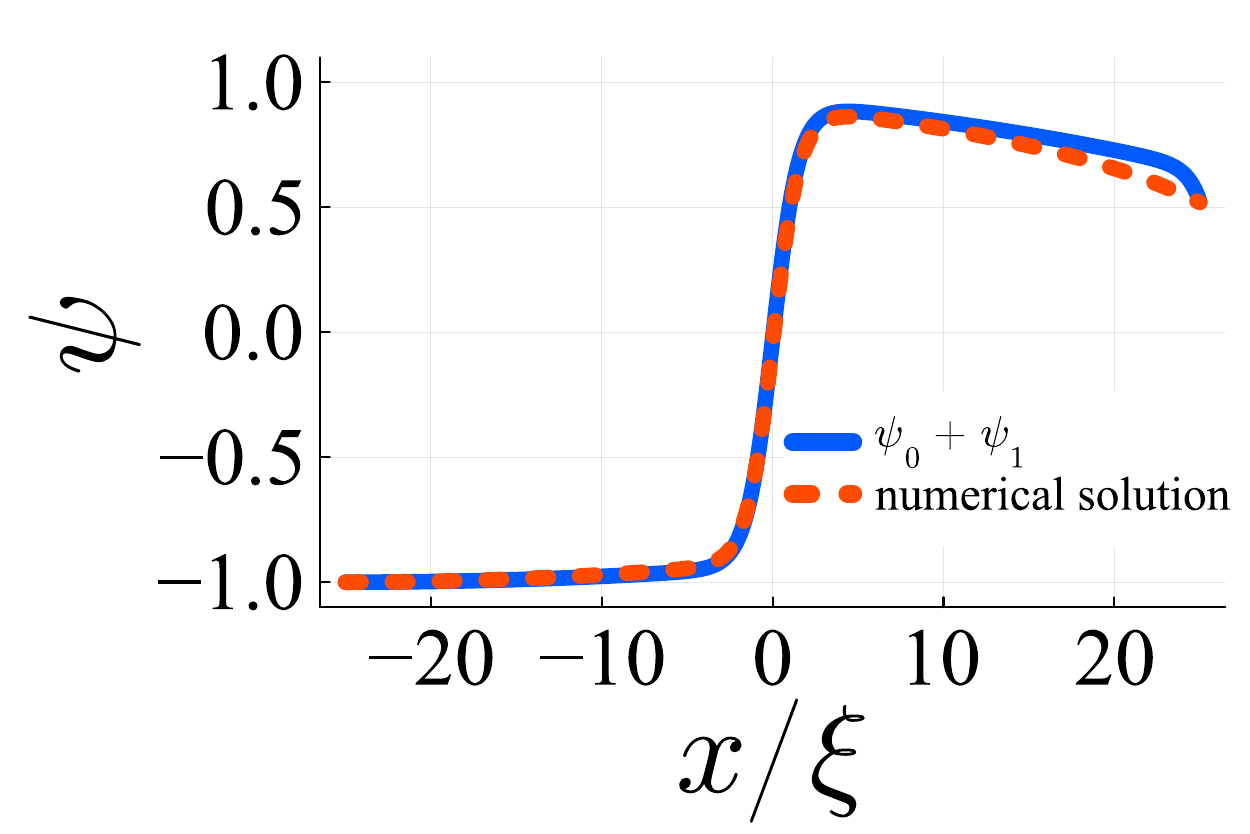}
\captionsetup{justification=raggedright,singlelinecheck=false}
\caption{Comparison between the approximated solution up to the first order, namely $\psi_0(x) + \psi_1(x)$ plotted in the blue solid curve, and the snapshot at the time $t=0$ in the time-evolution obtained from the numerical calculation we discuss in Sec.~\ref{grad_spin_num} plotted in the red dashed curve.} 
\label{fig: comparison-psi1} 
\end{figure}

We note the consistency between the numerical and the analytical results. Figure~\ref{fig: comparison-psi1} shows that the function $\psi_0(x) + \psi_1(x)$ is a good approximation to the numerical result $\psi(x,t)$.

\begin{figure}
\vspace{5cm}
\includegraphics[width=70mm]{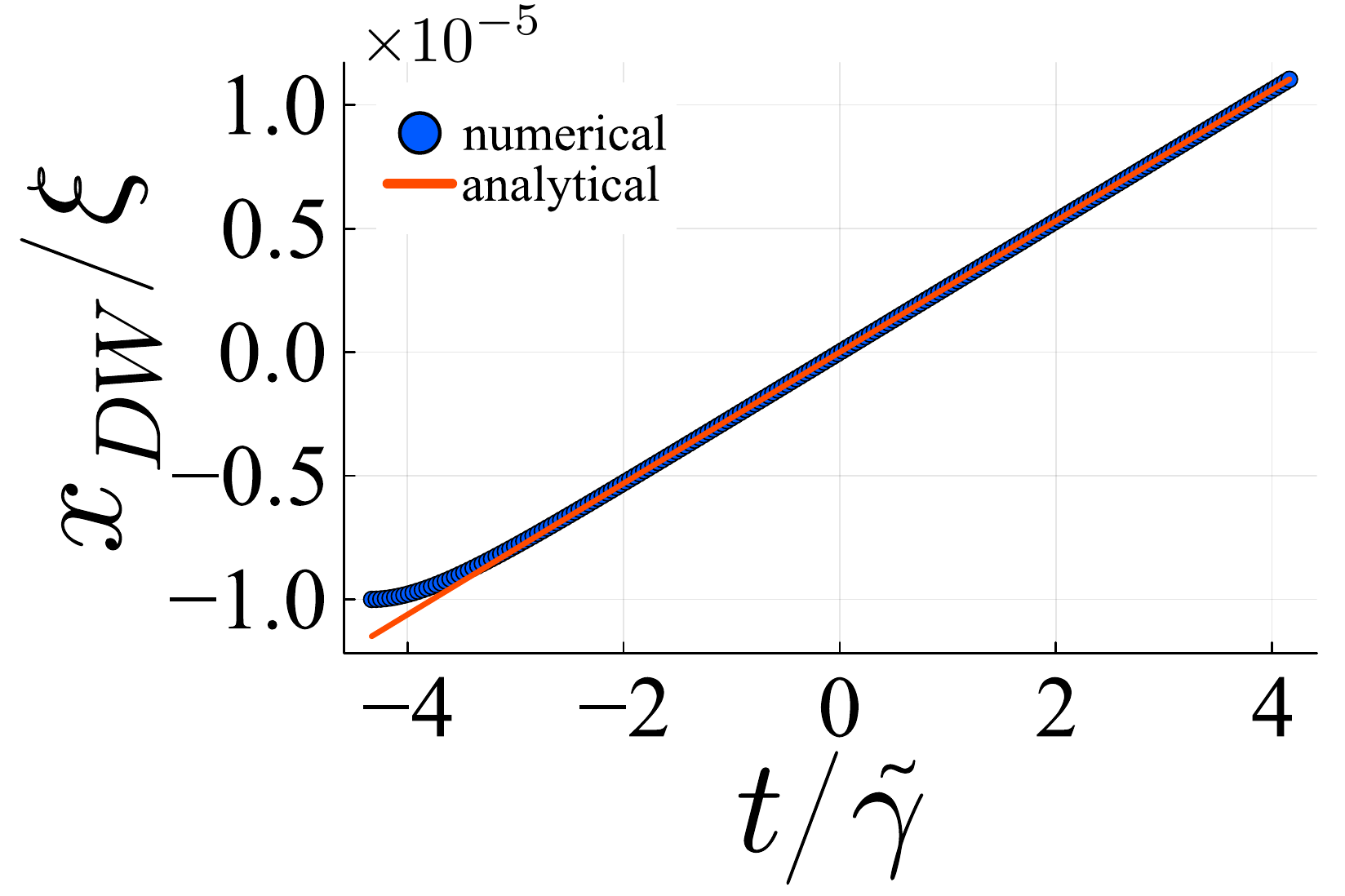}
\captionsetup{justification=raggedright,singlelinecheck=false}
\caption{Time-dependence of the coordinate of the domain wall. The solid red line and the blue curve represent the analytical and numerical result under the parameters $x_{\textrm{R}}/\xi=25$, $\sigma^{\textrm{sp}}_{\textrm{s}}/\sigma^{\textrm{sp}}_{\textrm{n}}=\tau^{\textrm{sp}}_{\textrm{s}}/\tau^{\textrm{sp}}_{\textrm{n}}=2$, $\xi^2/(\sigma_{\textrm{n}}\tau_{\textrm{n}})=1/1000$, $\xi^2/(\sigma_{\textrm{n}}\tilde{\gamma})=10$, $T=0.99T_{\textrm{c}}(\mu^2_{\textrm{R}})$ and $\mu_{\rm R}/\Delta_0 = 0.0010$. The ratio between the analytically obtained velocity $v_{\rm ana}$ and the numerically obtained velocity $v_{\rm m}$ is $v_{\rm sim} / v_{\rm ana}=0.9994$ (see Supplemental Material \cite{sm_ratio_vana_vnum} for $\mu_{\rm R}$-dependency of the ratio.).}
\label{fig: x-t-curve-spin}
\end{figure}

Figure~\ref{fig: x-t-curve-spin} compares the time dependence of the domain wall coordinate obtained from the numerical calculation with the result of the linear analysis (Eq.~\eqref{vvsdmu2dx}) under the same parameters. Here, the highly spin-accumulated boundary condition is $\mu_{\rm R}/\Delta_0 = 0.0010$, and the other parameters are equal to the values we adopted in FIG.~\ref{figspingradient}. We used a different value of $\mu_{\rm R}/\Delta_0$ to validate the linearity of $\psi(x,t)$ and $\mu^2(x,t)$ with respect to the velocity of the domain wall. For the dependence of this ratio on $\mu_{\rm R}/\Delta_0$, see Supplemental Material~\cite{sm_ratio_vana_vnum}.
\subsection{Remarks}
We considered systems with temperature and spin accumulation gradients and found that the two cases are similarly understood. In both cases, the domain wall moves toward the region of smaller order-parameter amplitude, namely, the higher temperature or the larger absolute value of spin accumulation.

We note the spin current $j_{\textrm{spin}}$, which is given in the form of 
\begin{equation} \label{jspin}
  j_{\textrm{spin}}(x,t) \equiv - \sigma^{\textrm{sp}}(x,t)\frac{\partial \mu(x,t)}{\partial x},
\end{equation}
is not a convenient quantity for describing the interaction between the domain wall and the spin density. 

When the sign of $\mu(x,t)$ is changed ($\mu \rightarrow -\mu$), the spin current defined above is also directed to the opposite. However, the dynamics of $\psi(x,t)$ and accordingly that of the domain wall remain unchanged under the sign reversal of $\mu(x,t)$. This is because $\mu(x,t)$ enters the TDGL equation Eq.~\eqref{TDGLspin} through $\mu^2(x,t)$.  
Therefore, we can conclude that the spin current given by Eq.~\eqref{jspin} does not interact with the domain wall; instead, the gradient of $\mu^2$ is essential.

\section{MOTION OF SINGLE VORTEX UNDER TEMPERATURE GRADIENT} \label{sec4vortex}

In this section, we extend the analytical methods developed for the
domain wall analysis in Sec.~\ref{sec2gradT} to an isolated vortex under
a temperature gradient.
The goal of this section is to derive an explicit expression for the velocity of the vortex. As we will see later, this expression closely corresponds to the one obtained in the domain wall case, and its sign indicates that the vortex moves toward the higher-temperature region.

We begin by redefining the model. We consider a three-dimensional superconductor and choose the axis of the isolated vortex to be along the $z$-axis, assuming that all physical quantities are independent of the $z$-coordinate. The system extends infinitely along the $z$-direction, and in the $xy$-plane, it is defined in the square region $-L/2 \leq x \leq +L/2$, $-L/2 \leq y \leq +L/2$. The vortex lies at $(x,y)=(0,0)$. We assume that the system length $L$ is much larger than both the coherence length and the magnetic penetration length. 

We set heat baths at the boundaries $x = \pm L/2$ with different temperatures $T_{\textrm{L}}$ at $x = -L/2$ and $T_{\textrm{R}}$ at $x = +L/2$. These temperatures satisfy the relation $T_{\textrm{L}} < T_{\textrm{R}} < T_{\textrm{c}}$. 

We treat the planes at $y=\pm \frac{L}{2}$ as boundaries with vacuum. 
A heat current flows in the negative $x$-direction through the superconductor.

\subsection{Model}
A complex condensate wavefunction
\begin{equation}
  \Psi(\bm{r},t) = f(\bm{r},t)e^{i\chi(\bm{r},t)},
\end{equation} 
is the order parameter. Here real functions $f$ and $\chi$ are the dimensionless amplitude and the phase of the order parameter, respectively.  It is convenient to convert the scalar and vector potentials $\Phi(\bm{r},t)$ and $\bm{A}(\bm{r},t)$ into gauge-invariant forms
\begin{subequations} 
\begin{align}
P(\bm{r},t) &\equiv \Phi(\bm{r},t) + \frac{\hbar}{e^{*}} \frac{\partial \chi(\bm{r},t)}{\partial t}  \\
\bm{Q}(\bm{r},t) &\equiv \bm{A}(\bm{r},t) - \frac{\hbar}{e^{*}} \bm{\nabla} \chi(\bm{r},t) 
\end{align}
\end{subequations}
with $e^{*}$ being the charge of a Cooper pair.

The following set of equations describes the vortex dynamics in the presence of the temperature gradient:
\begin{widetext}
\begin{subequations} \label{eqsystems} 
    \begin{align} 
\tilde{\gamma}\frac{\partial f'(\bm{r}',t')}{\partial t'}=\left[\frac{1}{\kappa^2}\nabla'^2 - \bm{Q}'^2(\bm{r}',t') - \frac{\tau'(\bm{r}',t')-1}{1-\tau'_{\textrm{L}}}\right]f'(\bm{r}',t')-f'^3(\bm{r}',t') \label{eqsystemsa} \\
    \tilde{\gamma} P'(\bm{r}',t') f'^2(\bm{r}',t')  +\frac{1}{\kappa} \bm{\nabla}' \cdot \left(f'^2(\bm{r}',t') \bm{Q}'(\bm{r}',t')\right)=0 \label{eqsystemsb} \\
    \bm{\nabla}' \times \left(\bm{\nabla}' \times \bm{Q}'(\bm{r}',t')\right) = \sigma'_{\textrm{n}}\left(-\frac{1}{\kappa}\grad' P'(\bm{r}',t') -\frac{\partial \bm{Q}'(\bm{r}',t')}{\partial t'}\right) -f'^2(\bm{r}',t')\bm{Q}'(\bm{r}',t') \label{eqsystemsc} \\
    C'\frac{\partial \tau'(\bm{r}',t')}{\partial t'} = \bm{\nabla}' \cdot \left[\left\{1+\left(k-1\right)f'^2(\bm{r}',t')\right\} \grad' \tau'(\bm{r}',t')\right]. \label{eqsystemsd}
    \end{align}
\end{subequations}
\end{widetext}

The prime symbol ($'$) indicates a dimensionless variable.  For the non-dimensionalization of these equations and the details of the boundary conditions, see the Supplemental Material. In particular, we choose the magnetic penetration length
$\lambda \equiv \sqrt{\frac{m^{*} \beta T_\textrm{c}}{\mu_0 e^{*2} \alpha_0 (T_\textrm{c} - T_\textrm{L})}}$
as the unit of length, where $\mu_0$ denotes the magnetic constant (=vacuum permeability), and $m^{*}$ the mass of a Cooper pair.  We define the Ginzburg–Landau parameter as $\kappa = \lambda / \xi$. Equations~\eqref{eqsystemsa}--\eqref{eqsystemsd} represent, respectively, the real part of the TDGL equation, its imaginary part (which describes the charge continuity equation), Ampère’s law, and the thermal diffusion equation. We solve this system of equations under the following boundary conditions:
\begin{widetext}
\begin{subequations} \label{vortexbcs}
    \begin{align}
    
    \left[\left(\frac{1}{\kappa}\grad' - i\bm{Q}'(\bm{r}',t')\right)f(\bm{r}',t')\right]\cdot \bm{n}&=0  \\
    \color{black}
    \bm{Q}'(\bm{r}',t') \cdot \bm{n} &= 0 \label{bcsforvortexheat4main}\\
    \left( \grad' P'(\bm{r}',t') + \frac{\partial \bm{Q}'(\bm{r}',t')}{\partial t'} \right) \cdot \bm{n} &= 0 \label{bcsforvortexheat5main}\\
    \tau'(-L'/2,y',t') &= \tau'_{\textrm{L}} \\
    \tau'(+L'/2,y',t') &= \tau'_{\textrm{R}} \\
    \left. \frac{\partial}{\partial y'}\tau'(x',y',t') \right|_{y'=\pm L'/2} &= 0. \label{bctauy}
    \end{align}
\end{subequations}
\end{widetext}
Hereafter, we omit the primes for simplicity.

The quantity $\sigma_{\textrm{n}}$ is the electrical conductivity. For simplicity, we ignore thermoelectric effects in both Ampère’s law and the thermal diffusion equation. In the boundary conditions~\eqref{bcsforvortexheat4main} and~\eqref{bcsforvortexheat5main}, the vector $\bm{n}$ represents the outward-pointing normal vector on the boundary.

\subsection{Linearlized equations}
We expand the quantities with respect to the vortex velocity $\bm{v}$ as
\begin{subequations} \label{vexpansions}
    \begin{align}
    f(\bm{r},t) &= f_0(\bm{r}-\bm{v}t)+f_1(\bm{r}-\bm{v}t)+\mathcal{O}(v^2) \\
    \bm{Q}(\bm{r},t) &= \bm{Q}_0(\bm{r}-\bm{v}t)+\bm{Q}_1(\bm{r}-\bm{v}t)+\mathcal{O}(v^2)  \\
    P(\bm{r},t) &= P_0(\bm{r}-\bm{v}t)+P_1(\bm{r}-\bm{v}t)+\mathcal{O}(v^2) \notag
\\& = P_1(\bm{r}-\bm{v}t)+\mathcal{O}(v^2) 
    \\
    \tau(\bm{r},t) &= \tau_0(\bm{r}-\bm{v}t)+\tau_1(\bm{r}-\bm{v}t)+\mathcal{O}(v^2) \notag
    \\& = \tau_{\textrm{L}} +\tau_1(\bm{r}-\bm{v}t)+\mathcal{O}(v^2).
    \end{align}
\end{subequations}
The subscripts 0 or 1 denote the order of the velocity $v$, as used in Eqs.~\eqref{expofpsi} and \eqref{expoftau}. 

In equilibrium where the vortex remains stationary, we can set $P_0 =0$ and $\tau_0=\tau_{\textrm{L}}$. 

We substitute Eqs.~\eqref{vexpansions} into Eqs.~\eqref{eqsystems} and obtain the expressions for the 0th and the 1st order equations. The 0th order equations yield the well-known equilibrium Ginzburg–Landau equations:
\begin{subequations} \label{vortexeqeq}
    \begin{align} \label{vortexeqeqa}
\left[\frac{1}{\kappa^2}\laplacian - Q_0^2(\bm{r}) +1\right]f_0(\bm{r})-f_0^3(\bm{r})=0, \\
\label{eqeqc} \bm{\nabla} \times \left(\bm{\nabla} \times \bm{Q}_0(\bm{r})\right) = -f_0^2(\bm{r})\bm{Q}_0(\bm{r}).
    \end{align}
\end{subequations}
The thermal diffusion equation~\eqref{eqsystemsd} has no contribution from the 0th order. In the following analysis, we use the fact that the solutions $f_0(\bm{r})$ and $\bm{Q}_0(\bm{r})$ obtained from these equations take the form $f_0(\bm{r}) = f_0(r)$ with $r = \lvert \bm{r} \rvert$, and $\bm{Q}_0(\bm{r}) = Q_0(r) \bm{e}_{\theta}$, where $\bm{e}_{\theta}$ denotes the tangential unit vector in two-dimensional polar coordinates. This expression for $\bm{Q}_0(\bm{r})$ directly yields the relation $\div \bm{Q}_0(\bm{r}) =0$.

We obtain the 1st order equations as follows: 
\begin{widetext}
\begin{subequations} \label{vortexeq1st}
    \begin{align} 
\left[\frac{1}{\kappa^2} \laplacian - Q_0^2(\bm{r}) + 1 - 3f_0^2(\bm{r})\right] f_1(\bm{r}) - 2f_0(\bm{r})\bm{Q}_0(\bm{r})\cdot\bm{Q}_1(\bm{r}) = -\tilde{\gamma} \bm{v} \cdot \grad f_0(\bm{r}) + \frac{\tau_1(\bm{r})}{1-\tau_{\textrm{L}}}f_0(\bm{r}) \label{vortexeq1sta} \\
\left[\laplacian - \frac{\tilde{\gamma} \kappa^2}{\sigma_{\textrm{n}}}f_0^2(\bm{r})\right]P_1(\bm{r})=0 \label{vortexeq1stb} \\
  \bm{\nabla} \times \left(\bm{\nabla} \times \bm{Q}_1(\bm{r})\right)+f_0^2(\bm{r})\bm{Q}_1(\bm{r})+2f_0(\bm{r})f_1(\bm{r})\bm{Q}_0(\bm{r}) = \sigma_{\textrm{n}}\left(-\frac{1}{\kappa}\grad  P_1(\bm{r}) + \left( \bm{v} \cdot \bm{\nabla} \right) \bm{Q}_0(\bm{r})\right) \label{vortexeq1stc} \\
  \bm{\nabla} \cdot \left[\left\{1+\left(k-1\right)f_0^2(\bm{r})\right\} \grad  \tau_1(\bm{r})\right] \label{vortexeq1std} = 0.
    \end{align}
\end{subequations}
\end{widetext}
Here, we focus on the time domain defined by \begin{equation} |\bm{v}t| \ll \xi, \ \lambda \ (\ll L), \label{eq: condition on vt vortex} \end{equation} which allows us to consider the vortex in a region sufficiently far from the boundaries (cf. Eq.~\eqref{eq: condition on vt}). We can first solve Eqs.~\eqref{vortexeq1stb} for $P_1(\bm{r})$ and \eqref{vortexeq1std} for $\tau_1(\bm{r})$. The solutions $P_1(\bm{r})$ and $\tau_1(\bm{r})$ then appear in the inhomogeneous terms in Eqs.~\eqref{vortexeq1sta} and \eqref{vortexeq1stc}. Equation~\eqref{vortexeq1std} is equivalent to the divergence-free condition of the heat flow $\bm{q}$ in the first order with respect to $\bm{v}$, which is given by
\begin{equation} \label{vortexqdefinition}
  \bm{q} = -\left[1+\left(k-1\right)f_0^2(\bm{r})\right] \grad  \tau_1(\bm{r}).
\end{equation}

\subsection{Zero-mode solutions}
Equations~\eqref{vortexeq1sta} and \eqref{vortexeq1stc} are the linear inhomogeneous differential equations for $f_1(\bm{r})$ and $\bm{Q}_1(\bm{r})$. It is known to be helpful to obtain the solutions of the corresponding homogeneous equations~\cite{GK1975,Dorsey92}. We refer to these solutions as the zero-mode solutions. 

Due to the translational symmetry of the equilibrium Ginzburg--Landau equations, $f_0(\bm{r} + \bm{d})$ and $\bm{Q}_0(\bm{r} + \bm{d})$ are also the solutions for Eqs.~\eqref{vortexeqeq}, where $\bm{d}$ is an arbitrary infinitesimal displacement vector within the $xy$  plane.  We expand these translated solutions as follows:
\begin{subequations}
    \begin{align}
    f_0(\bm{r}+\bm{d}) &= f_0(\bm{r}) + \left( \bm{d} \cdot \bm{\nabla} \right) f_0(\bm{r}) + \mathcal{O}(\lvert \bm{d} \rvert^2), \\
    \bm{Q}_0(\bm{r}+\bm{d}) &= \bm{Q}_0(\bm{r}) + \left( \bm{d} \cdot \bm{\nabla} \right) \bm{Q}_0(\bm{r}) + \mathcal{O}(\lvert \bm{d} \rvert^2).
    \end{align}
\end{subequations}

We substitute these expansions into the equilibrium Ginzburg--Landau equations~\eqref{vortexeqeq} for $f_0(\bm{r} + \bm{d})$ and $\bm{Q}_0(\bm{r} + \bm{d})$, and extract the first-order terms in $\bm{d}$. This procedure yields the following results:
\begin{subequations} \label{zeromodes}
    \begin{align}
    &\left[ \frac{1}{\kappa^2} \laplacian 
    - Q_0^2(\bm{r}) + 1 - 3f_0^2(\bm{r}) \right] 
    f_{\textrm{d}}(\bm{r}) \notag\\
    &\quad - 2f_0(\bm{r}) \bm{Q}_0(\bm{r}) 
    \cdot \bm{Q}_{\textrm{d}}(\bm{r}) = 0, 
    \label{zeromodesa} \\
    &\bm{\nabla} \times 
    \left( \bm{\nabla} \times \bm{Q}_{\textrm{d}}(\bm{r}) \right) 
    + f_0^2(\bm{r}) \bm{Q}_{\textrm{d}}(\bm{r}) \notag\\
    &\quad + 2f_0(\bm{r}) f_{\textrm{d}}(\bm{r}) 
    \bm{Q}_0(\bm{r}) = \bm{0}. 
    \label{zeromodesb}
    \end{align}
\end{subequations}
We introduce the following notation:
\begin{subequations} \label{deffdQd}
    \begin{align}
    f_{\textrm{d}}(\bm{r}) &\equiv \left( \bm{d} \cdot \bm{\nabla} \right) f_0(\bm{r}), \\
    \bm{Q}_{\textrm{d}}(\bm{r}) &\equiv \left( \bm{d} \cdot \bm{\nabla} \right) \bm{Q}_0(\bm{r}).
    \end{align}
\end{subequations}

We thus identify Eqs.~\eqref{zeromodes} as the homogeneous equations corresponding to the inhomogeneous equations~\eqref{vortexeq1sta} and \eqref{vortexeq1stc} for $f_1(\bm{r})$ and $\bm{Q}_1(\bm{r})$. Consequently, we recognize $f_{\textrm{d}}(\bm{r})$ and $\bm{Q}_{\textrm{d}}(\bm{r})$ as the desired zero-mode solutions.

We then rewrite equations~\eqref{vortexeq1sta} and \eqref{vortexeq1stc} by utilizing the zero-mode solutions. First, we multiply the Eq.~\eqref{vortexeq1sta} by $f_{\textrm{d}}(\bm{r})$ from the left and integrate it over the entire system. We note that each term in the integrands on both sides is spatially localized around the vortex core $(r \lesssim \xi$, $\lambda)$. As a result, the dominant contribution to the integral comes from the vicinity of the vortex core, and that from the far region is negligible. We can change the integration region from the entire system to a circular area with radius $R_{\textrm{c}}$. We assume $\xi$, $\lambda \ll R_{\textrm{c}} \ll L$, so that we can identify $f_0(\bm{r})$ and $\bm{Q}_0(\bm{r})$ to be the solutions for Eq.~\eqref{vortexeqeq} in an infinite system, not a finite system, under the boundary conditions $f_0(r) \rightarrow 1$, $Q_0(r) \rightarrow 0$ when $r \rightarrow \infty$. In what follows, we denote this integration region by $\textrm{C}$. This procedure yields the following equation:
\begin{align} \label{intfd}
  \int_{\textrm{C}} \textrm{d}\bm{r} \ \bigg[
&2f_0(\bm{r})f_1(\bm{r})\bm{Q}_0(\bm{r})\cdot\bm{Q}_\textrm{d}(\bm{r}) \notag
    \\ &- 2f_0(\bm{r})f_{\textrm{d}}(\bm{r})\bm{Q}_0(\bm{r})\cdot\bm{Q}_1(\bm{r}) \notag\\
    &+ f_{\textrm{d}}(\bm{r})\frac{1}{\kappa^2} \laplacian f_1(\bm{r}) 
    - f_1(\bm{r})\frac{1}{\kappa^2} \laplacian f_{\textrm{d}}(\bm{r})
  \bigg] \notag
  \\ &= 
  \int_{\textrm{C}} \textrm{d}\bm{r}  \ \left[-\tilde{\gamma} f_{\textrm{d}}(\bm{r}) \bm{v} \cdot \grad f_0(\bm{r}) + \frac{\tau_1(\bm{r})}{1-\tau_{\textrm{L}}}f_{\textrm{d}}(\bm{r})f_0(\bm{r}) \right].
\end{align}
Here, we used integration by parts and Eq.~\eqref{zeromodesa}. 

Similarly, we take the inner product of Eq.~\eqref{vortexeq1stc} with 
$\bm{Q}_{\textrm{d}}(\bm{r})$ from the left and integrate it over the entire space. We have
\begin{align} \label{intQd}
  \int_{\textrm{C}} \textrm{d}\bm{r} \ \bigg[
    &2f_0(\bm{r})f_1(\bm{r})\bm{Q}_0(\bm{r}) \cdot \bm{Q}_\textrm{d}(\bm{r}) \notag
    \\&- 2f_0(\bm{r})f_{\textrm{d}}(\bm{r})\bm{Q}_0(\bm{r}) \cdot \bm{Q}_1(\bm{r}) \notag
    \\&+ \bm{Q}_{\textrm{d}}(\bm{r}) \cdot \left( \bm{\nabla} \times \left( 
        \bm{\nabla} \times \bm{Q}_1(\bm{r}) \right) \right) \notag
    \\&- \bm{Q}_1(\bm{r}) \cdot \left( \bm{\nabla} \times \left( 
        \bm{\nabla} \times \bm{Q}_{\textrm{d}}(\bm{r}) \right) \right)
  \bigg] \notag \\
  &= \int_{\textrm{C}}\textrm{d}\bm{r}\; 
  \bm{Q}_{\textrm{d}}(\bm{r}) \cdot \left( \sigma_{\textrm{n}} \bm{E}_{1}(\bm{r}) \right),
\end{align}
where the symbol $\bm{E}_1(\bm{r})$ is defined by
\begin{equation}
 \bm{E}_{1}(\bm{r}) \equiv \left(-\frac{1}{\kappa}\grad  P_1(\bm{r}) + \left( \bm{v} \cdot \bm{\nabla} \right) \bm{Q}_0(\bm{r})\right).
\end{equation}

Equations~\eqref{intfd} and \eqref{intQd} give 
\begin{align} \label{BTf}
  \int_{\textrm{C}} \textrm{d}\bm{r}\; \Big(
    \sigma_{\textrm{n}} \bm{Q}_{\textrm{d}}(\bm{r}) \cdot \bm{E}_1(\bm{r})
    + \tilde{\gamma} f_{\textrm{d}}(\bm{r}) \left(\bm{v} \cdot \bm{\nabla}\right)f_0(\bm{r}) 
    \notag \\
    - \frac{\tau_1(\bm{r})}{1 - \tau_{\textrm{L}}} f_{\textrm{d}}(\bm{r}) f_0(\bm{r})
  \Big) = \textrm{B.T.},
\end{align}
where
\begin{align}
  \textrm{B.T.} &= \int_{\textrm{C}}\textrm{d}\bm{r}\; \Big[
    \bm{Q}_{\textrm{d}}(\bm{r}) \cdot \left(\bm{\nabla} \times \left(\bm{\nabla} \times \bm{Q}_1(\bm{r})\right)\right)
    \notag \\
    &\hspace{2.3cm}
    - \bm{Q}_1(\bm{r}) \cdot \left(\bm{\nabla} \times \left(\bm{\nabla} \times \bm{Q}_{\textrm{d}}(\bm{r})\right)\right)
  \Big]
  \notag \\
  &\quad - \int_{\textrm{C}}\textrm{d}\bm{r}\;\Big[
    f_{\textrm{d}}(\bm{r}) \frac{1}{\kappa^2} \laplacian f_0(\bm{r})
    - f_0(\bm{r}) \frac{1}{\kappa^2} \laplacian f_{\textrm{d}}(\bm{r})
  \Big]
\end{align}
are the boundary terms, which are negligible in the limit $L \rightarrow \infty$. Extracting the explicit $\bm{v}$-dependency of the integral term with $\bm{E}_1$ in Eq.~\eqref{BTf}, we have
\begin{equation} \label{intE1}
  \int_{\textrm{C}}\textrm{d}\bm{r}\;\sigma_{\textrm{n}} \bm{Q}_{\textrm{d}}(\bm{r}) \cdot \bm{E}_1(\bm{r}) = \int_{\textrm{C}} \textrm{d}\bm{r}\;  \sigma_{\textrm{n}} \bm{Q}_{\textrm{d}}(\bm{r}) \cdot  \left( \bm{v} \cdot \bm{\nabla} \right) \bm{Q}_0(\bm{r}).
\end{equation}
There is no contribution from the term with $\grad P_1$ in $\bm{E}_1$ because it becomes a negligible boundary term. 

\subsection{Expression for $\bm{v}$}
Since we can choose $\bm{d}$ arbitrarily, we let it parallel to $\bm{v}$ and $\bm{q}$, and take all these vectors in the $x$-direction, assuming the motion of the vortex is limited to only along the $x$-direction. In this case, we write $\bm{d} \cdot \bm{\nabla} = d \partial/\partial x$ and $\bm{v} \cdot \bm{\nabla} = v \, \partial/\partial x$, where $\bm{d} = d \bm{e}_x$ and $\bm{v} = v \bm{e}_x$, with $\bm{e}_x$ denoting the unit vector in the $x$-direction. The constant $d$ cancels out from all terms. Eventually, from Eqs.~\eqref{BTf} and \eqref{intE1}, we arrive at the following expression (cf. Eq.~\eqref{vvsqtransport}):
\begin{align} \label{vortexvelocity}
v &= \frac{
  \displaystyle  \frac{1}{2(1 - \tau_{\textrm{L}})} \int_{\textrm{C}} \textrm{d}\bm{r}\; 
   \tau_1(\bm{r}) \frac{\partial}{\partial x} f_0^2(\bm{r})
}{
  \displaystyle \int_{\textrm{C}} \textrm{d}\bm{r}\;\left[
    \tilde{\gamma} \left( \frac{\partial f_0(\bm{r})}{\partial x} \right)^2 
    + \sigma_{\textrm{n}} \left| \frac{\partial}{\partial x} \bm{Q}_0(\bm{r}) \right|^2
  \right]} \notag
  \\&=\frac{
  \displaystyle \frac{1}{2(1 - \tau_{\textrm{L}})} \int_{\textrm{C}} \textrm{d}\bm{r}\;
  \frac{\partial}{\partial x}\tau_1(\bm{r})(f_0^2(R_{\textrm{c}})-f_0^2(\bm{r}))
}{
  \displaystyle \int_{\textrm{C}} \textrm{d}\bm{r}\;\left[
    \tilde{\gamma} \left( \frac{\partial f_0(\bm{r})}{\partial x} \right)^2 
    + \sigma_{\textrm{n}} \left| \frac{\partial}{\partial x} \bm{Q}_0(\bm{r}) \right|^2
  \right]}.
\end{align}
Equation~\eqref{vortexqdefinition} and the assumption $\bm{q} =q\bm{e}_x$, $q<0$ lead
\begin{align}
  \frac{\partial}{\partial x}\tau_1(\bm{r}) = \frac{-q}{1+\left(k-1\right)f_0^2(\bm{r})}>0.
\end{align}

This inequality implies that $v>0$, indicating that the vortex moves
toward the higher-temperature region. The resulting direction of the vortex
motion is the same as that obtained analytically for the domain wall
model, and the two cases share the same physical picture. The only
additional feature in the vortex case is the term proportional to
$\sigma_{\textrm{n}}$, which can be interpreted as a Joule-heating
contribution to the viscous force.
\subsection{numerical simulation} \label{subsec_num_vor_heat}
\begin{widetext}
\begin{figure*}
\includegraphics[width=180mm]{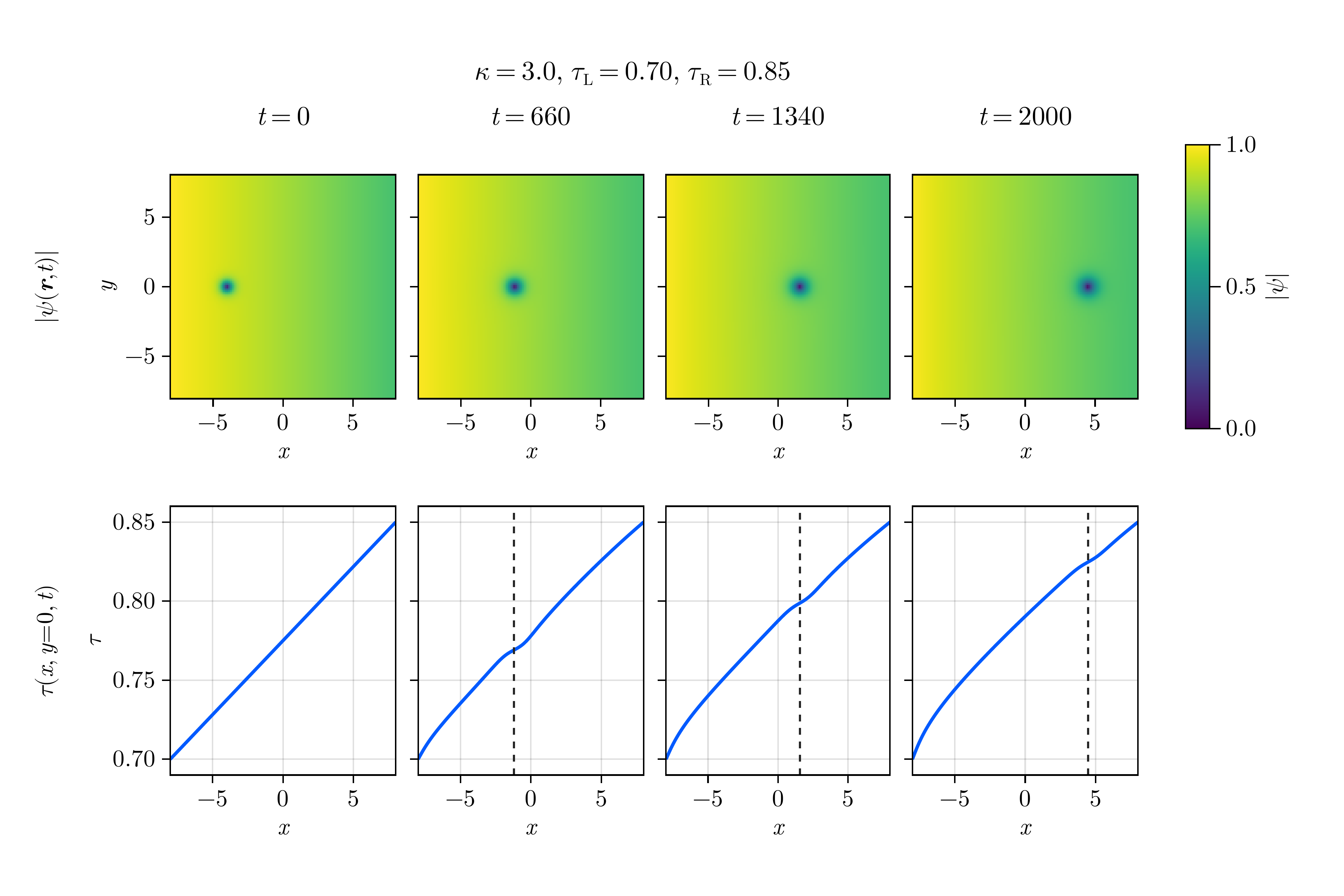}
\captionsetup{justification=raggedright,singlelinecheck=false}
\caption{Snapshots of the numerical solutions for an isolated vortex under a temperature gradient. The upper and lower rows show the spatial distributions of the order parameter amplitude $|\psi(x,y,t)|$ and the temperature $\tau(x,y,t)$, respectively. The vertical dashed line indicates the $x$-coordinate of the vortex center. The unit of length is $\lambda$. These results show that the vortex moves toward the higher-temperature boundary while the local temperature gradient reduced within the vortex core.}
\label{num_vortex_heat} 
\end{figure*}
\end{widetext}

Figure~\ref{num_vortex_heat} shows the numerical results for the single vortex dynamics governed by these equations. We set the parameters to $L=16$, $\kappa=3$, $\tau_{\textrm{L}}=0.70$, $\tau_{\textrm{R}}=0.85$, $C=30.0$, and $k=0.050$, with no external magnetic field. The upper panels show color plots of the order parameter amplitude, while the lower panels show the $x$-dependence of the temperature field along the cross section through the vortex center ($y=0$). As the initial condition at $t=0$, we put an isolated vortex at $(x,y)=(-4,0)$ and impose a linear temperature profile along the $x$-direction, which is uniform along the $y$-direction. The resulting time evolution reveals that the vortex moves toward the higher temperature boundary. We also find that the temperature gradient locally flattens within the vortex core region, reflecting the enhanced thermal conductivity in the normal core. These findings are highly consistent with the domain wall dynamics shown in Fig.~\ref{figthermalgeadient}.
\color{black}

\section{MOTION OF A SINGLE VORTEX UNDER
SPIN DENSITY GRADIENT} \label{sec_vortex_spin}
\subsection{analytical calculation}
We use a similar model to analyze the dynamics of an isolated vortex in the presence of a spin-accumulation gradient. We write down the dimensionless system of equations as
\begin{widetext}
\begin{subequations} \label{eqsystemsspin} 
    \begin{align} 
\tilde{\gamma}\frac{\partial f'(\bm{r}',t')}{\partial t'}=\left[\frac{1}{\kappa^2}\nabla'^2 - \bm{Q}'^2(\bm{r}',t') - \left(\frac{\tau'}{\tau'_{\textrm{c}}\left( \mu'^2\right)}-1\right)\left(1-\frac{\tau'}{\tau'_{\textrm{c}}\left( 0\right)}\right)^{-1}\right]f'(\bm{r}',t')-f'^3(\bm{r}',t') \label{eqsystemssspina} \\
    \tilde{\gamma} P'(\bm{r}',t') f'^2(\bm{r}',t')  +\frac{1}{\kappa} \bm{\nabla}' \cdot \left(f'^2(\bm{r}',t') \bm{Q}'(\bm{r}',t')\right)=0 \label{eqsystemsspinb} \\
    \bm{\nabla}' \times \left(\bm{\nabla}' \times \bm{Q}'(\bm{r}',t')\right) = \sigma'_{\textrm{n}}\left(-\frac{1}{\kappa}\grad' P'(\bm{r}',t') -\frac{\partial \bm{Q}'(\bm{r}',t')}{\partial t'}\right) -f'^2(\bm{r}',t')\bm{Q}'(\bm{r}',t') \label{eqsystemsspinc} \\
    a'_1\frac{\partial \mu'^2(\bm{r}',t')}{\partial t'} = \bm{\nabla}'\cdot \left[\sigma'^{\textrm{sp}}(\bm{r}',t') \grad' \mu'^2(\bm{r}',t')\right]-\frac{\sigma'^{\textrm{sp}}(\bm{r}',t')}{2\mu'^2(\bm{r}',t')}\left(\grad' \mu'^2(\bm{r}',t')\right)^2-a'_2\frac{2\mu'^2(\bm{r}',t')}{\tau'^{\textrm{sp}}(\bm{r}',t')}, \label{eqsystemsspind}
    \end{align}
\end{subequations}
\end{widetext}
where 
\begin{subequations}
    \begin{align}
    \displaystyle
    \sigma'^{\textrm{sp}}(\bm{r}',t') = \sigma'^{\textrm{sp}}_{\textrm{n}}+\left( \sigma'^{\textrm{sp}}_{\textrm{s}}-\sigma'^{\textrm{sp}}_{\textrm{n}}\right)f'^2(\bm{r}',t')  \\
    \displaystyle
   \tau'^{\rm sp}(\bm{r}',t') = \tau'^{\rm sp}_{\textrm{n}}+\left( \tau'^{\rm sp}_{\textrm{s}}-\tau'^{\rm sp}_{\textrm{n}}\right)f'^2(\bm{r}',t') \\ \displaystyle
   a'_1 = \alpha \lambda^2 / (\hbar \sigma^{\textrm{sp}}_{\textrm{n}}) \\ \displaystyle
   a'_2 = \lambda^2 / (\sigma^{\textrm{sp}}_{\textrm{n} } \tau^{\textrm{sp}}_{\textrm{n}}). 
    \end{align}
\end{subequations}
The prime symbol ($'$) indicates a dimensionless variable.  For the non-dimensionalization of these equations and the details of the boundary conditions, see the Supplemental Material. The boundary conditions are
\begin{subequations} \label{vortexbcsspin}
   \begin{align}
   
    \left[\left(\frac{1}{\kappa}\grad' - i\bm{Q}'(\bm{r}',t')\right)f(\bm{r}',t')\right]\cdot \bm{n}&=0 
    \color{black} \\
    \bm{Q}'(\bm{r}',t') \cdot \bm{n} &= 0 \\
    \left( \grad' P'(\bm{r}',t') + \frac{\partial \bm{Q}'(\bm{r}',t')}{\partial t'} \right) \cdot \bm{n} &= 0 \\
    \mu'^2(-L'/2,y',t') &= 0 \\
    \mu'^2(+L'/2,y',t') &= \mu'^2_{\textrm{R}} \\
    \left. \frac{\partial}{\partial y'}\mu'^2(x',y',t') \right|_{y'=\pm L'/2} &= 0. \label{bcmu2y}
    \end{align}
\end{subequations}
Hereafter, we omit the primes for simplicity. Expanding Eqs.~\eqref{eqsystemsspin} with respect to $\bm{v}$, we obtain the first-order relations
\begin{subequations}\label{vortexeq1stspin}
\begin{align}
  \MoveEqLeft{\left[
      \frac{1}{\kappa^2} \laplacian
      - Q_0^2(\bm{r}) + 1- 3f_0^2(\bm{r})
    \right] f_1(\bm{r})} \notag \\
    &- 2f_0(\bm{r}) \bm{Q}_0(\bm{r}) \cdot \bm{Q}_1(\bm{r}) \notag
 \\ & = - \tilde{\gamma} \bm{v} \cdot \nabla f_0(\bm{r}) 
    -b\mu_1^2(\bm{r}) f_0(\bm{r})
    \label{vortexeq1stspina} \\
    0 &= \div \left[\sigma^{\textrm{sp}}_0(\bm{r}) \grad  \mu_1^2(\bm{r})\right] \notag \\ 
    & -\frac{\sigma^{\textrm{sp}}_0(\bm{r})}{2\mu_1^2(\bm{r})}\left(\grad \mu_1^2(\bm{r})\right)^2 -\frac{2\mu_1^2(\bm{r})}{\tau^{\textrm{sp}}_0(\bm{r})}. \tag{\ref{vortexeq1stspin}d}
    \label{vortexeq1stspind}
\end{align}
\end{subequations}
Equations~\eqref{vortexeq1stspina} and \eqref{vortexeq1stspind} are derived from the real part of the TDGL equation~\eqref{eqsystemssspina} and the spin diffusion equation~\eqref{eqsystemsspind}, respectively. 
Here, we have introduced the following quantities:
\begin{subequations}
    \begin{align}
    \sigma^{\textrm{sp}}_0(\bm{r}) &= \sigma^{\textrm{sp}}_{\textrm{n}}+\left( \sigma^{\textrm{sp}}_{\textrm{s}}-\sigma^{\textrm{sp}}_{\textrm{n}}\right)f_0^2(\bm{r}), \\
    \tau^{\rm sp}_0(\bm{r}) &= \tau^{\rm sp}_{\textrm{n}}+\left( \tau^{\rm sp}_{\textrm{s}}-\tau^{\rm sp}_{\textrm{n}}\right)f_0^2(\bm{r}).
    \end{align}
\end{subequations}
Equations~\eqref{eqsystemsb} and ~\eqref{eqsystemsc} require no correction. After the same calculation as we have performed in Sec.~\ref{sec4vortex}, we have the linear relation between $\bm{v}$ and $\mu_1^2$, which corresponds to Eq.~\eqref{BTf} as
\begin{align} \label{BTfspin}
  \int_{\textrm{C}} \textrm{d}\bm{r}\; \Big(
    \sigma_{\textrm{n}} \bm{Q}_{\textrm{d}}(\bm{r}) \cdot \bm{E}_1(\bm{r})
    + \tilde{\gamma} f_{\textrm{d}}(\bm{r}) \left(\bm{v} \cdot \bm{\nabla}\right)f_0(\bm{r}) 
    \notag \\
    +b\mu_1^2(\bm{r}) f_{\textrm{d}}(\bm{r}) f_0(\bm{r})
  \Big) = \textrm{B.T.}.
\end{align}
This relation directly yields \begin{align} \label{vortexvelocityspin}
v &= \frac{
  \displaystyle -\frac{b}{2}\int_{\textrm{C}} \textrm{d}\bm{r}\;\mu_1^2(\bm{r})\frac{\partial}{\partial x} f_0^2(\bm{r})
}{
  \displaystyle \int_{\textrm{C}} \textrm{d}\bm{r}\;\left[
    \tilde{\gamma} \left( \frac{\partial f_0(\bm{r})}{\partial x} \right)^2 
    + \sigma_{\textrm{n}} \left| \frac{\partial}{\partial x} \bm{Q}_0(\bm{r}) \right|^2
  \right]} \notag
  \\&=\frac{
  \displaystyle -\frac{b}{2}\int_{\textrm{C}} \textrm{d}\bm{r}\;\frac{\displaystyle \partial \mu_1^2(\bm{r})}{\partial x}(f_0^2(R_{\textrm{c}})-f_0^2(\bm{r}))
}{
  \displaystyle \int_{\textrm{C}} \textrm{d}\bm{r} \;\left[
    \tilde{\gamma} \left( \frac{\partial f_0(\bm{r})}{\partial x} \right)^2 
    + \sigma_{\textrm{n}} \left| \frac{\partial}{\partial x} \bm{Q}_0(\bm{r}) \right|^2
  \right]}.
\end{align}
See Eqs.~\eqref{vvsdmu2dx} and \eqref{vortexvelocity} for the corresponding expressions. We can identify the sign of the velocity to be positive (see Appendix \ref{appendixmu12deriv}), which shows that the vortex moves to the region with a larger modulus of the spin accumulation.

\subsection{numerical simulation} \label{subsec_num_vor_spin}
\begin{widetext}
\begin{figure*}
\includegraphics[width=180mm]{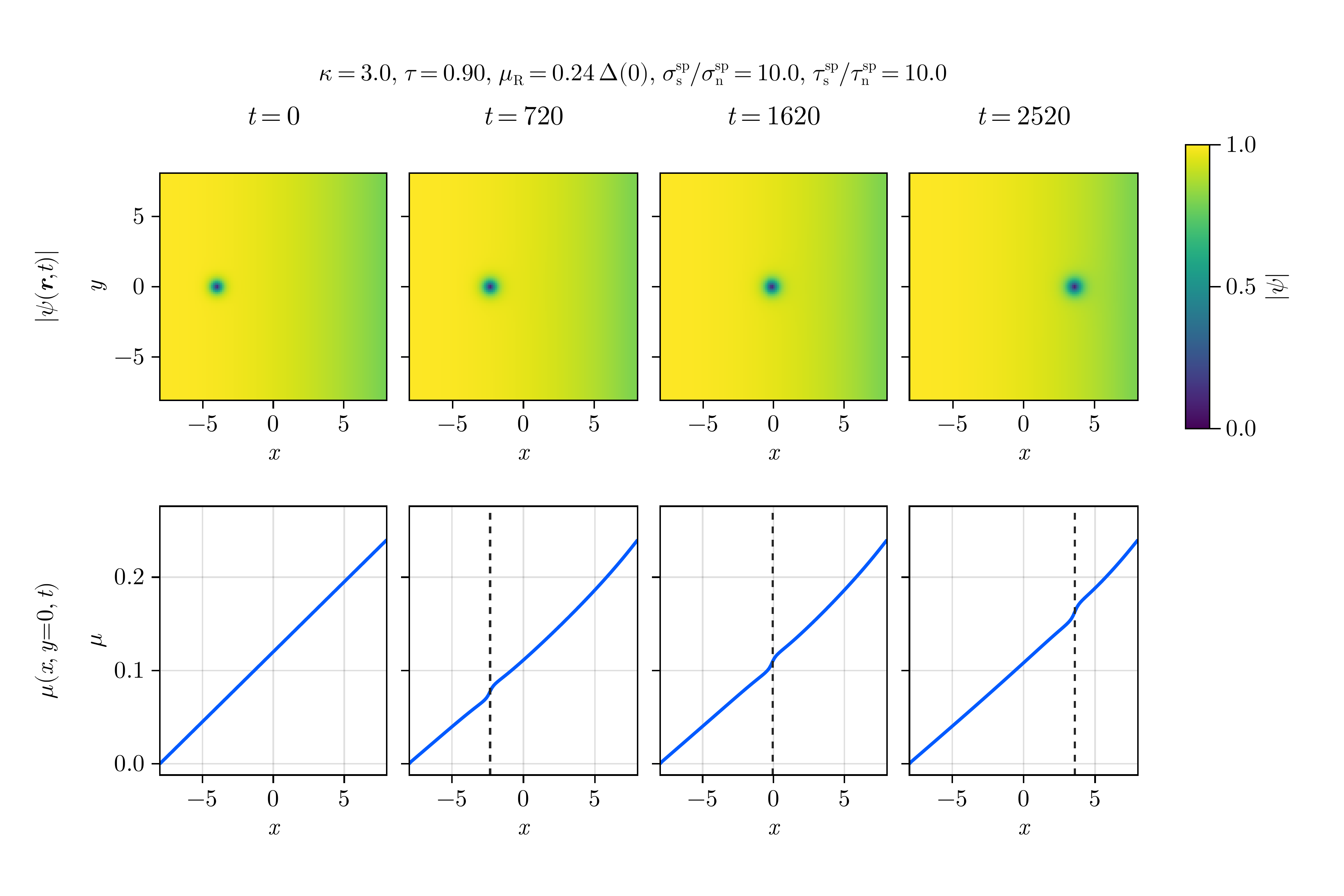}
\captionsetup{justification=raggedright,singlelinecheck=false}
\caption{Snapshots of the numerical solutions for an isolated vortex under a spin-density gradient. The upper and lower rows show the spatial distributions of the order parameter amplitude $|\psi(x,y,t)|$ and the spin accumulation $\mu(x,y,t)$, respectively. The vertical dashed line indicates the $x$-coordinate of the vortex center.  The unit of length is $\lambda$, and the unit of spin accumulation is the bulk superconducting gap at zero temperature. These results show that the vortex moves toward the boundary with larger $\mu$ while the local spin accumulation gradient becomes steeper within the vortex core.}
\label{num_vortex_spin} 
\end{figure*}
\end{widetext}

Figure~\ref{num_vortex_spin} shows the numerical results for the single vortex dynamics governed by these equations. We set the parameters to $L=16$, $\kappa=3$, $\tau=0.90$, $\mu_{\textrm{R}}=0.240\Delta(0)$, $\sigma^{\textrm{sp}}_{\textrm{s}}/\sigma^{\textrm{sp}}_{\textrm{n}}=10.0$, and $\tau^{\textrm{sp}}_{\textrm{s}}/\tau^{\textrm{sp}}_{\textrm{n}}=10.0$, with no external magnetic field. The upper panels show color plots of the order parameter amplitude, while the lower panels show the $x$-dependence of the local spin accumulation along the cross section through the vortex center ($y=0$). As the initial condition at $t=0$, we put an isolated vortex at $(x,y)=(-4,0)$ and impose a linear $\mu$ profile along the $x$-direction, which is uniform along the $y$-direction. The resulting time evolution reveals that the vortex moves toward the higher spin accumulation boundary. We also find that the spin accumulation gradient locally steepens within the vortex core. These findings are highly consistent with the domain wall dynamics shown in Fig.~\ref{figspingradient}.
\color{black}

\section{Discussion}
\label{sec3.5discussion}
Our result in Sec.~\ref{sec2gradT} shows that the thermal force on the domain wall in the present study is a direct consequence of the local momentum balance relation in the TDGL equation, where the spatial variation of temperature is only taken into account in the coefficient of the linear term of the order parameter. In Ref.~\cite{Kopnin1975}, the transport entropy per vortex per unit length along the vortex line is written in sec. 4 in Ref.~\cite{Kopnin1975} as
\begin{equation}
S_d=\frac{N_0\xi^2}{T}\int{\rm d}\bm{r}\;(|\Delta_\infty|^2-|\Delta(\bm{r})|^2).    
\end{equation}
Here, $N_0$ is the density of one-particle states at the Fermi level in the normal state.
Combining this expression and Eq.~\eqref{Fthstephen} and comparing it with   Eq.~\eqref{eq: Fth-2}, we find that Eq.~\eqref{eq: Fth-2} is a thermal force with the sign opposite to that of  Stephen's thermal force Eq.~\eqref{Fthstephen}.  

Our result in Sec.~\ref{sec4vortex} implies that an isolated vortex
moves toward the hotter region in a temperature gradient. This finding is in marked contrast to earlier studies
\cite{OtterSolomon66, SOLOMON1968293}, in which vortices were considered
to move toward the colder region.

\color{black}

The experiments~\cite{OtterSolomon66,SOLOMON1968293} were performed in
states with a finite density of vortices.

This discrepancy between Refs.~\cite{OtterSolomon66,SOLOMON1968293}
and the present study implies that dilute and dense vortex states may
behave differently in the presence of a temperature gradient.

In the present study, we find that the driving force (momentum flow into the region containing the domain wall) is negligible when the system size is much larger than the coherence length $\xi$. 

The driving force may become significant when the vortex
density is finite, and the resulting vortex motion can differ from that
of an isolated vortex.
In Ref.~\cite{Stephen66}, Stephen attributed the motion of vortices
toward the colder region to weaker intervortex repulsion in the colder
region [see the last seven lines on p.~802 of Ref.~\cite{Stephen66}].
This effect is absent in the isolated vortex. 

The directional transition of vortex motion between dilute and dense
vortex states in the presence of a temperature gradient remains an issue
for future study.
  
A particularly important field for further application of the results in Sec.~\ref{sec3gradmu} and Sec.~\ref{sec_vortex_spin} is spintronics in superconductors. Among these, the vortex spin Hall effect, which was recently examined in Refs.~\cite{KimSeKwon2018,vargunin2019flux,taira2021spin,adachi2024timedependent}, provides specific evidence of the critical role of moving vortices in the spin transport in superconductors. These studies have been carried out in a region under a higher magnetic field and thus are complementary to the present work.

A further difference between our work and 
Refs.~\cite{Dorsey92,Sergeev2010} lies in how heat flow 
is modeled. 
In Refs.~\cite{Dorsey92,Sergeev2010}, heat is carried by 
the moving vortex---through the transport energy 
$U_\phi$~\cite{Dorsey92} or the core entropy 
$S_d^{\rm core}$~\cite{Sergeev2010}---and no thermal 
diffusion equation for $T(\bm{r},t)$ is included. In contrast, our model treats heat flow as quasiparticle diffusion, 
$\bm{q} = -\kappa[|\psi|^2]\nabla T$, and solves for 
$T(\bm{r},t)$ directly. 
Near $T_c$, vortex-mediated contributions vanish 
($U_\phi \propto (1-T/T_c)\to 0$ and $S_d^{\rm core}\to 0$), whereas $\kappa_n$ remains finite. Within the near-$T_c$ regime where the TDGL
description applies, quasiparticle diffusion is therefore the relevant
heat-flow channel in our model.

The domain wall model contains no electromagnetic field, yet it gives the same direction of motion toward
the hotter region. This supports our interpretation that, within the
present TDGL-diffusion model, the hotter-region motion originates from
the order-parameter dynamics rather than from vortex-mediated
electromagnetic heat transport.
\color{black}

We compare the results on the domain wall with those on the vortex. The model describing the dynamics of the domain wall does not include electromagnetic fields, and the order parameter is taken to be real. The existence of the domain wall is ensured by the boundary conditions at $x=x_{\textrm{L,R}}$, where the order parameter takes opposite signs. In contrast, the model for the vortex dynamics includes the electromagnetic potentials and requires solving not only the TDGL equation and the heat (or spin) diffusion equation, but also Ampère’s law. The electromagnetic field couples to the phase of the order parameter, and the TDGL equation is decomposed into its real and imaginary parts. Despite these differences in the structure of the models, we have demonstrated that
the same analytical strategy based on the momentum balance relation can be applied to both systems. The resulting velocity expressions for the domain wall and the vortex have closely analogous structures. The additional dissipative term from the normal current appears only in the vortex model. Since this term has the same sign as the dissipative contribution from the TDGL relaxation term, it enhances the damping but does not change the direction of motion.

In this sense, the domain wall analysis captures the same mechanism
that appears in the vortex calculation, while the two systems remain physically distinct.

We can also consider the significance of addressing
the domain wall problem rather than vortices. A prime
example is the application to the FFLO state, where
we can treat the periodic array of domain walls as a
periodic extension of the system discussed in our study. Recently, the possible FFLO state in zero magnetic field has been discussed~\cite{SumitaFFLO}, and the dynamics of these states under inhomogeneous temperature or spin-density gradients is a significant future extension.

Finally, we comment on the possible effect of the vortex Hall angle on the vortex dynamics. Distinct families of superconductors, such as electron-doped and hole-doped high-$T_{\textrm{c}}$ cuprates, can exhibit different vortex Hall responses, reflecting differences in their underlying material properties \cite{blatter,armitage2010progress}. In the present study, as a first step toward clarifying the fundamental driving mechanisms, we restricted our analysis to the longitudinal motion along the applied gradients. We therefore used the dissipative TDGL equation with a real relaxation coefficient and did not include the vortex Hall effect.

The vortex Hall effect can be incorporated into the TDGL framework by allowing the relaxation coefficient to be complex \cite{Dorsey92, KC, Sugai}. This modification would introduce a transverse component in addition to the longitudinal motion discussed in this work. Since the vortex Hall response depends on the material, the direction and magnitude of this transverse component would depend on the vortex Hall angle of the specific compound. Importantly, this extension does not change the basic structure of the momentum-balance approach \cite{KC, Sugai}, although it would require a separate quantitative analysis. A detailed treatment of vortex motion including the Hall angle is therefore left as an important future problem.
\color{black}

\color{black}

\section{Conclusion} \label{sec5conc}
To summarize, we studied the dynamics of topological defects, namely a
domain wall and a vortex, in type-II superconductors under a temperature
gradient or a spin-accumulation gradient. We solved the TDGL equation
together with the thermal or spin diffusion equation, whose coefficients
vary spatially and temporally depending on the order parameter.
Our numerical and analytical results show that the domain wall moves
toward the hotter boundary in the presence of a temperature gradient,
and toward the region with the larger magnitude of spin accumulation in
the presence of a spin-accumulation gradient. We have obtained analytical
expressions for the velocities of the domain wall and the vortex as
functions of the heat flow or the spin-accumulation profile.

The domain wall velocity expressions were quantitatively validated by
comparison with numerical simulations. For the vortex case, the
simulations confirm the predicted direction of motion under both
temperature and spin-accumulation gradients.
The motion of topological defects in the presence of temperature or
spin-accumulation gradients can be understood as a consequence of
dynamics that reduce the loss of condensation energy.

We also derived expressions for the forces on the domain wall, including
those due to the temperature and spin-accumulation gradients. The force
due to the temperature gradient has the same form but with the opposite
sign to that derived by Stephen for a superconducting vortex. The
discrepancy between our result and Stephen's argument suggests that
isolated defects and dense vortex states can behave differently under a
temperature gradient.

This theoretical study sheds light on the long-standing problem of
superconducting vortex motion by combining a tractable
domain wall prototype with explicit single-vortex calculations.
The present framework also provides a theoretical basis for exploring
efficient methods to control superconducting vortices and FFLO domain
walls via heat flow or spin current.
\begin{acknowledgments}
We thank J.-i. Ohe for his helpful comments on our manuscript. T.K. was financially supported by World-Leading
Innovative Graduate Study Program of Advanced Basic Science Course (WINGS-ABC) of the University of Tokyo and JST SPRING, Grant Number JPMJSP2108. This research was supported by JSPS
KAKENHI Grants (No. JP22H01941 and No. JP21H01799).
\end{acknowledgments}

\appendix
\section{Expression for $\psi_1$ and $\tau_1$} \label{appendix1storder}
\begin{figure*}
\includegraphics[width=120mm]{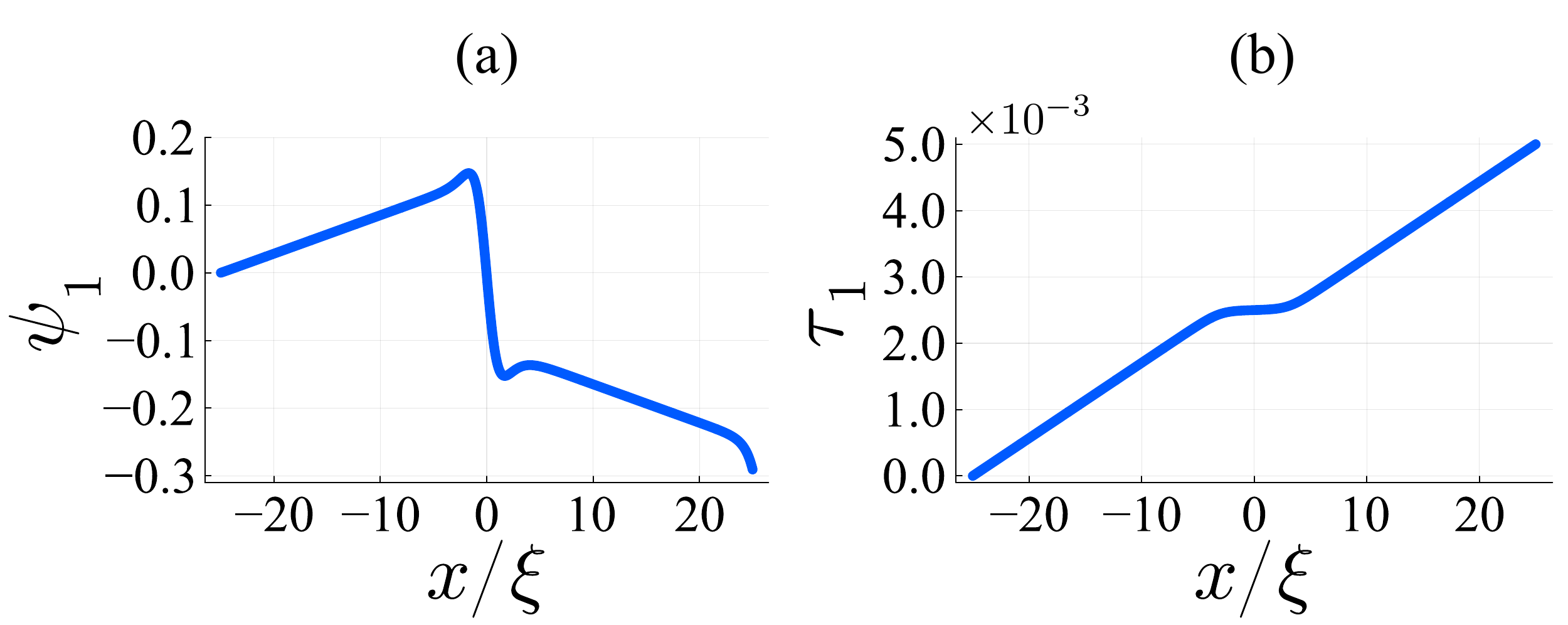}
\captionsetup{justification=raggedright,singlelinecheck=false}
\caption{(a) Profile of $\psi_1(x)$ in Eq.~\eqref{eq: psi1-app} with $R(x)$ given by Eq.~\eqref{eq: Rx}, and (b) that of $\tau_1(x)$ in Eq.~\eqref{eq: tau-1-app}. Parameters are the same as those described in Sec.~\ref{gradtnum}.}
\label{fig1st} 
\end{figure*}

Here, we present the expressions for $\tau_1(x)$ and $\psi_1(x)$. The shapes of these functions are shown in FIG.~\ref{fig1st}. 
Equation~\eqref{inttau1-without-c} yields the explicit expression for $\tau_1(x)$ as follows:
\begin{align}
\tau_1(x)&=\frac{(\tau_{\rm R}-\tau_{\rm L})}{2}\left(\frac{
F(x;k)}{F(x_{\rm R};k)}+1\right),
\label{eq: tau-1-app}
\end{align}
with 
\begin{align}
F(x;k)&:=
\int_{0}^{x} \frac{\text{d}x'}{1+(k-1)\psi_0^2(x')}\notag\\
&=\frac{\sqrt{2}\xi}{1-k}\tanh\frac{x}{\sqrt{2}\xi}\notag\\
&+\frac{\sqrt{2}\xi k}{(1-k)\frac32}\ln\left|\frac{(1-k)^\frac12\tanh\frac{x}{\sqrt{2}\xi}-1}{(1-k)^\frac12\tanh\frac{x}{\sqrt{2}\xi}+1}\right|.

\end{align}

Next, we solve Eq.~\eqref{tdgl1storder} to find $\psi_1(x)$. The expression for $\psi_1(x)$ is less explicit.  
The function
\begin{align}
y_1(x) &\equiv \xi \frac{\text{d}\psi_0(x)}{\text{d}x}  \notag
\\ &= \frac{1}{\sqrt{2} \cosh ^2 \left(x/\sqrt{2}\xi\right)}
\end{align}
is one of the fundamental solutions to 
the homogeneous differential equation
\begin{equation} \label{Lpsi1=0}
  L_0(x)\psi_1(x) = 0.
\end{equation}

 Another fundamental solution to Eq.~\eqref{Lpsi1=0} is given by
\begin{align}
  y_2(x) &= \frac{y_1(x)}{\xi} \int_{x_{\textrm{L}}}^{x} \frac{\text{d}x'}{y_1^2(x')} \notag
  \\&= \frac{Y(x)-Y(x_{\textrm{L}})}{\cosh ^2 \left(x/\sqrt{2}\xi\right)},
\end{align}
where 
\begin{equation}
  Y(x) \equiv \frac{\sinh ^2 \left(\sqrt{2} x/\xi\right)}{2}+\frac{\sinh ^2 \left(2\sqrt{2} x/\xi\right)}{16}+\frac{3x}{4 \sqrt{2}}.
\end{equation}
The integral depends on $x$, and thus, the linear independence between $y_1(x)$ and $y_2(x)$ is obvious. With the use of the method of ^^ ^^ variation of constant", we obtain the analytical solution to $\hat{L}(x)\psi_1(x)=R(x)$ with 
\begin{equation}
  R(x) \equiv \tilde{\gamma}v \frac{\text{d}\psi_0(x)}{\text{d}x} - \frac{\psi_0(x)\tau_1(x)}{1- \tau_{\textrm{L}}}
  \label{eq: Rx}
\end{equation}
as
\begin{align}
  \MoveEqLeft{\psi_1(x) = C_1 y_1(x) + C_2 y_2(x)} \notag \\& - y_1(x)\int_{x_{\textrm{L}}}^{x} \text{d}x' \frac{R(x')y_2(x')}{W(x')} \notag \\& + y_2(x) \int_{x_{\textrm{L}}}^{x} \text{d}x' \frac{R(x')y_1(x')}{W(x')}.
  \label{eq: psi1-app}
\end{align}
Here 
\begin{equation}
  C_1 = 0
\end{equation}
and
\begin{align}
C_2 &= \frac{1}{y_2(x_{\textrm{R}})}\left(\psi_1(x_{\textrm{R}})+y_1(x_{\textrm{R}})\int_{x_{\textrm{L}}}^{x_{\textrm{R}}} \text{d}x' \frac{R(x')y_2(x')}{W(x')}\notag \right. \\& \left. \hspace{0.3cm}  -y_2(x_{\textrm{R}})\int_{x_{\textrm{L}}}^{x_{\textrm{R}}} \text{d}x' \frac{R(x')y_1(x')}{W(x')} \right),
\end{align}
are constants determined from the boundary conditions. The function 
\begin{equation}
  W(x) \equiv \xi \left(y_1(x)\frac{\text{d}y_2(x)}{\text{d}x}-\frac{\text{d}y_1(x)}{\text{d}x}y_2(x)\right) 
\end{equation}
is the Wronskian of $y_1(x)$ and $y_2(x)$.

\section{Proof of the inequality \eqref{mu2mon}} \label{appendixmu12deriv}
From Eq.~\eqref{spindiffusionmu2}, the inequality 
\begin{align}
\MoveEqLeft{
\frac{\text{d}}{\text{d}x}\left[\sigma^{\textrm{sp}}_0(x)\frac{\text{d} \mu^2_1(x)}{\text{d}x} \right]} \notag \\
  &=\frac{\sigma^{\textrm{sp}}_0(x)}{2\mu^2_1(x)}\left(\frac{\text{d}\mu^2_1(x)}{\text{d} x}\right)^2+\frac{2\mu^2_1(x)}{\tau^{\textrm{sp}}
_0(x)}>0 
\end{align}
holds, implying that the function $\sigma^{\textrm{sp}}_0(x)\frac{\text{d} \mu^2_1(x)}{\text{d}x}$ is a monotonically increasing function with respect to $x$. Thus 
\begin{equation}
  \sigma^{\textrm{sp}}_0(x_{\textrm{L}})\left. \frac{\text{d}\mu^2_1(x)}{\text{d}x} \right|_{x=x_\textrm{L}} \leq \sigma^{\textrm{sp}}_0(x)\frac{d \mu^2_1(x)}{\text{d}x}
\end{equation}
or equivalently
\begin{equation}
\frac{\sigma^{\textrm{sp}}_0(x_{\textrm{L}})}{\sigma^{\textrm{sp}}_0(x)}\left. \frac{\text{d}\mu^2_1(x)}{\text{d}x} \right|_{x=x_\textrm{L}} \leq\frac{d \mu^2_1(x)}{\text{d}x}.
\end{equation}
holds for any $x$ in the system. 
Taking into account $\mu^2_1(x) \geq 0$ and $\mu^2_1(x_\textrm{L})=0$, we obtain the inequality~\eqref{mu2mon}.

\color{black}

%% file: SM2.tex
\subsection{Consistency between analytical and numerical results}
We discuss the ratio between the analytically obtained velocity $v_{\textrm{ana}}$ and the numerically obtained velocity $v_{\textrm{sim}}$ discussed in Sec.~\ref{comp.gradT} and Sec.~\ref{compgradmu}. 

\begin{figure}[ht] 
    \centering
    \includegraphics[width=70mm]{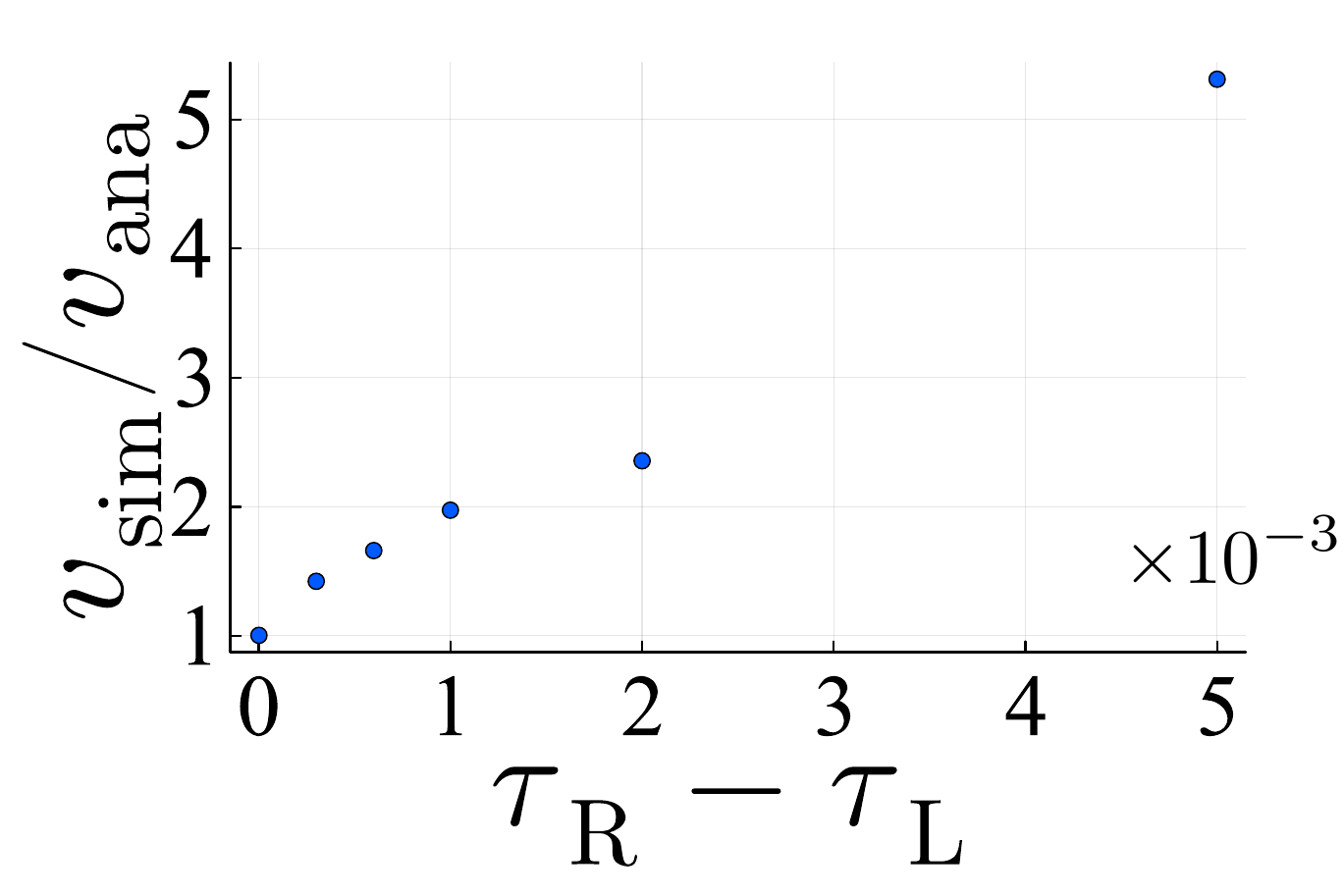}
    \caption{Temperature dependency of the ratio between the analytically obtained velocity $v_{\textrm{ana}}$ and the numerically obtained velocity $v_{\textrm{sim}}$.}
    \label{gradT}
\end{figure}

\vspace{1mm}
FIGURE~\ref{gradT} shows the temperature dependency of the ratio between $v_{\textrm{ana}}$ and $v_{\textrm{sim}}$. We fix $\tau_{\textrm{L}} = 0.990$, and vary $\tau_{\textrm{R}} = 0.995, 0.992, 0.991, 0.9906, 0.9903,0.990+10^{-6}$. To find $v_{\textrm{sim}}$, we follow the position of the domain wall in a specific region which includes the origin $x=0$. We perform the cubic spline interpolation to the discrete data to define $v_{\textrm{sim}}$ as the slope measured at the time when the domain wall passed through the origin. We evaluate $v_{\textrm{ana}}$ based on Eq.~\eqref{vvsqtransport}.

The ratio deviates from 1 in the region with large temperature differences. This deviation indicates that the linear analysis is not valid. On the other hand, as the temperature difference decreases, the ratio approaches 1, and the results of the linear analysis become consistent with those of the numerical calculation.

\begin{figure*}[ht]  
\centering
\includegraphics[width=70mm]{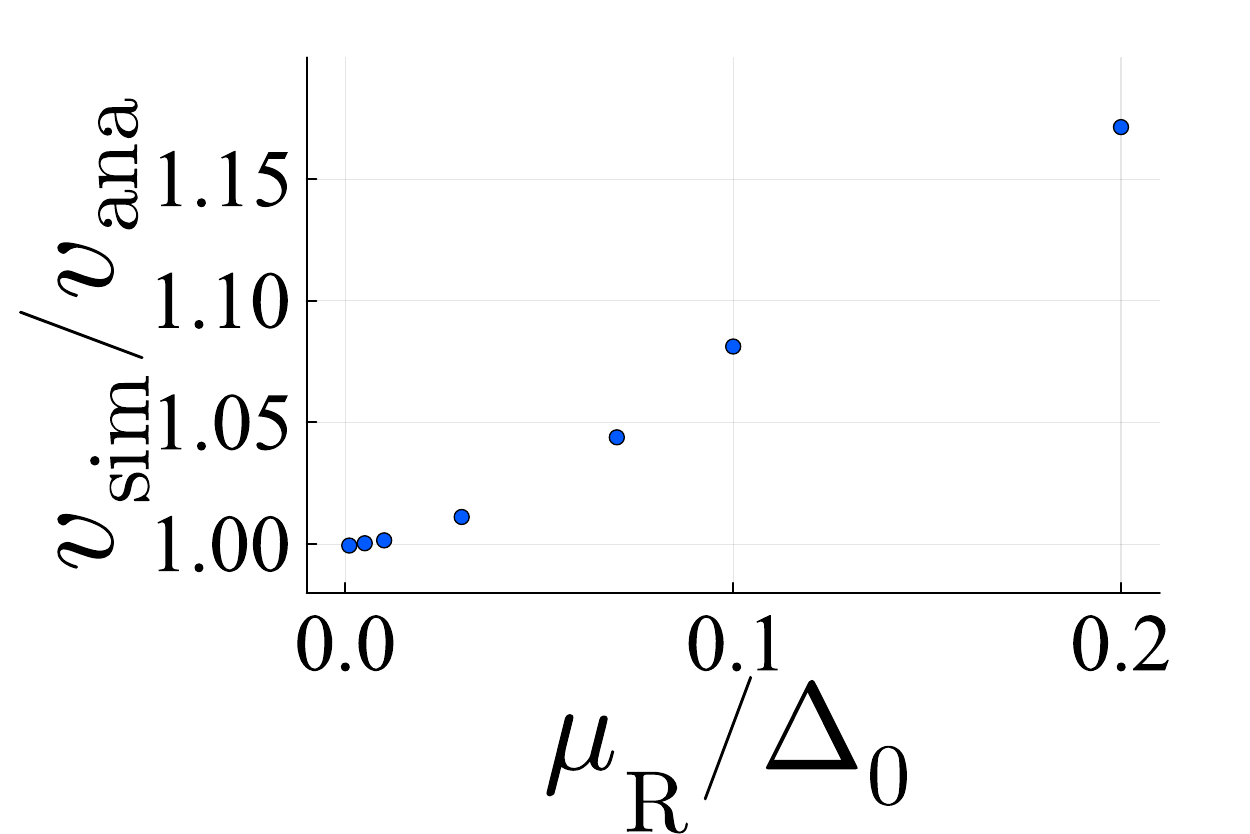}
\caption{$\mu_{\textrm{R}}$-dependency of the ratio between the analytically obtained velocity $v_{\textrm{ana}}$ and the numerically obtained velocity $v_{\textrm{sim}}$.}
\label{figspin}
\end{figure*}

FIGURE~\ref{figspin} shows the temperature dependency of the ratio between $v_{\textrm{ana}}$ and $v_{\textrm{sim}}$. We vary $\mu_{\textrm{R}}/\Delta_0 = 0.20,0.10,0.070,0.030,
    0.010,0.0050,0.0010$. We obtain $v_{\textrm{sim}}$ in the same manner as that for the inhomogeneous temperature case. We evaluate $v_{\textrm{ana}}$ based on Eq.~\eqref{vvsdmu2dx}. The function $\mu_1(x)$ which appears in Eq.~\eqref{vvsdmu2dx} is the solution of the linearized equation of Eq.~\eqref{SDeq}:
    \begin{equation} 
      \frac{\text{d}^2 \mu_{1}(x)}{\text{d}x^2} + \frac{1}{\sigma^{\textrm{sp}}_{0}(x)} \frac{\text{d}\sigma^{\textrm{sp}}_{0}(x)}{\text{d}x} \frac{\text{d} \mu_{1}(x)}{\text{d}x} - \frac{ \mu_{1}(x)}{\tau^{\textrm{sp}}_0(x)\sigma^{\textrm{sp}}_{0}(x)}=0
    \end{equation}
under the boundary conditions $\mu_{1}(x_{\textrm{L}})=0$, $\mu_{1}(x_{\textrm{R}})=\mu_{\textrm{R}}$. We rewrite the differential equation in terms of a new variable $z \equiv \tanh ^2{\displaystyle \left(\frac{x}{\sqrt{2}\xi}\right)}$. It becomes 
\begin{align}
    \frac{\text{d}^2\mu_1(z)}{\text{d}z^2} &+ \left[ \frac{1}{2z} - \frac{1}{1-z} + \frac{(\sigma^{\textrm{sp}}_{\textrm{r}}-1)}{1+(\sigma^{\textrm{sp}}_{\textrm{r}}-1)z} \right] \frac{\text{d}\mu_1(z)}{\text{d}z} \nonumber \\
    &- \frac{\xi^2}{2\tau_{\textrm{n}}\sigma^{\textrm{sp}}_{\textrm{n}}} \frac{\mu_1(z)}{z (1-z)^2\left\{1+(\tau_{\textrm{r}}-1)z\right\}\left\{1+(\sigma^{\textrm{sp}}_{\textrm{r}}-1)z\right\}} = 0.
\end{align}
where $\sigma^{\textrm{sp}}_{\textrm{r}} \equiv \sigma^{\textrm{sp}}_{\textrm{s}}/\sigma^{\textrm{sp}}_{\textrm{n}}$ and $\tau^{\textrm{sp}}_{\textrm{r}} \equiv \tau^{\textrm{sp}}_{\textrm{s}}/\tau^{\textrm{sp}}_{\textrm{n}}$. In the specific case where $\sigma^{\textrm{sp}}_{\textrm{r}} = \tau^{\textrm{sp}}_{\textrm{r}} \equiv s$, the equation has the following solution:
\begin{equation}
\mu_1(z) = c_1 \cosh\left(M(z)\right)+ c_2 \sinh\left(M(z)\right),
\end{equation}
where 
 \begin{equation}
  M(z) \equiv \frac{\xi }{s\sqrt{2 \sigma^{\textrm{sp}}_{\textrm{n}} \tau^{\textrm{sp}}_{\textrm{n}}}}\left\{ 2 \sqrt{s-1} \tan^{-1} \left[\sqrt{z(s-1)} \right]+\log\left( \displaystyle s \frac{\sqrt{z}+1}{\sqrt{z}-1}\right) \right\}.
\end{equation}
The symbols $c_1$ and $c_2$ are constants identified by the boundary conditions. 

The ratio deviates from 1 in the region with the large modulus of the spin accumulation. This deviation is attributed to the presence of non-linearity. When $\mu_{\textrm{R}}/\Delta_0$ is sufficiently small, the ratio approaches $1$, validating the results of the linear analysis.

\subsection{Equations and Boundary conditons for Sec.~\ref{sec4vortex}}
\subsubsection{Dimensionless Equations}

In this subsection, we summarize the system of equations introduced in Sec.~\ref{sec4vortex} and their non-dimensionalization. We describe the model using the complex order parameter
\begin{equation}
\Psi(\bm{r},t)=f(\bm{r},t)e^{i\chi(\bm{r},t)},
\end{equation}
the electromagnetic potentials
\begin{subequations} 
\begin{align}
P(\bm{r},t) &\equiv \Phi(\bm{r},t) + \frac{\hbar}{e^{*}} \frac{\partial \chi(\bm{r},t)}{\partial t}, \\
\bm{Q}(\bm{r},t) &\equiv \bm{A}(\bm{r},t) - \frac{\hbar}{e^{*}} \bm{\nabla} \chi(\bm{r},t),
\end{align}
\end{subequations}
and the local temperature
\begin{equation}
  T(\bm{r},t).
\end{equation}
We first write down the governing equations in their dimensional forms. The TDGL equation for the complex order parameter is
\begin{align}
\label{TDGL1}
\MoveEqLeft{\gamma\left(\frac{\partial}{\partial t}+\frac{ie^{*}}{\hbar}\Phi(\bm{r},t)\right)\Psi(\bm{r},t)} \notag
  \\&= \frac{\hbar^2}{2m^{*}}\left(\bm{\bm{\nabla}}-\frac{ie^{*}\bm{A}(\bm{r},t)}{\hbar}\right)^2\Psi(\bm{r},t)-\alpha_0 \frac{T(\bm{r},t)-T_\textrm{c}}{T_\textrm{c}} \Psi(\bm{r},t)-\beta \left| \Psi(\bm{r},t) \right|^2\Psi(\bm{r},t),
\end{align}
and Ampère's law is
\begin{equation}
  \bm{\bm{\nabla}} \times \left( \bm{\bm{\nabla}} \times \bm{Q}(\bm{r},t)\right) = \mu_0 \left( \bm{J}_{\textrm{n}}(\bm{r},t)+\bm{J}_{\textrm{s}}(\bm{r},t)\right).
\end{equation}
The supercurrent density is given by
\begin{align} \label{jsQ}
  \bm{J}_{\textrm{s}}(\bm{r},t)= -\frac{e^{*2}}{m^{*}} f^2(\bm{r},t) \bm{Q}(\bm{r},t),
\end{align} 
and the normal current density is
\begin{equation} \label{jnPQ}
  \bm{J}_{\textrm{n}}(\bm{r},t) = \sigma_{\textrm{n}}\left(- \grad P(\bm{r},t)-\frac{\partial \bm{Q}(\bm{r},t)}{\partial t}\right).
\end{equation}
Using the specific heat $C$ and the heat flux $\bm{q}(\bm{r},t)$, we write the heat diffusion equation as
\begin{equation} \label{TDeq}
  C\frac{\partial T(\bm{r},t)}{\partial t} + \div \bm{q}(\bm{r},t) = 0.
\end{equation}
Similar to the normal current density, we express the heat flux in terms of the temperature gradient:
\begin{equation} \label{SMdefofq}
  \bm{q}(\bm{r},t) = -K(\bm{r},t) \grad T(\bm{r},t).
\end{equation}
We assume that the thermal conductivity $K$ varies in space and time through the amplitude of the order parameter:
\begin{equation} \label{kappafirst}
K(\bm{r},t) = K_{\textrm{n}} + (K_{\textrm{s}}-K_{\textrm{n}})\frac{f^2(\bm{r},t)}{f^2_{\textrm{bulk}}},
\end{equation}
where
\begin{equation} \label{fbulk}
  f_{\textrm{bulk}} = \sqrt{-\frac{\alpha_0}{\beta}\frac{T_\textrm{L}-T_\textrm{c}}{T_\textrm{c}} }
\end{equation}
is the spatially uniform bulk solution of the equilibrium Ginzburg-Landau equation in the absence of a magnetic field.

Next, we summarize the non-dimensionalization of the physical quantities appearing in the above equations. In the following expressions, primed quantities represent dimensionless variables:
\begin{equation}
  \bm{r}=\lambda \bm{r}', \ L=\lambda L',
\end{equation}
\begin{equation}
  t=\frac{\hbar}{\alpha}t', \ \alpha = \alpha_0 \frac{T_\textrm{c}-T_\textrm{L}}{T_\textrm{c}}
\end{equation}
\begin{equation}
  \tilde{\gamma}=\frac{\gamma}{\hbar}
\end{equation}
\begin{equation}
f\left(\lambda\bm{r}',\frac{\hbar}{\alpha}t'\right)=  f_{\textrm{bulk}} f'(\bm{r},t) 
\end{equation}
\begin{equation} \label{dimlessP}
  P\left(\lambda\bm{r}',\frac{\hbar}{\alpha}t'\right)=\frac{e^{*}}{\alpha}P'(\bm{r}',t')
\end{equation}
\begin{equation}
\label{dimlessQ}
\bm{Q}\left(\lambda\bm{r}',\displaystyle\frac{\hbar}{\alpha}t'\right)=\sqrt{2}B_{\textrm{c}}\lambda \bm{Q}'(\bm{r}',t'), \ B_{\textrm{c}} =\alpha\sqrt{\frac{\mu_0}{\beta}}
\end{equation}
\begin{equation}
T\left(\lambda\bm{r}',\displaystyle\frac{\hbar}{\alpha}t'\right)= T_{\textrm{c}} \tau'(\bm{r}',t') 
\end{equation}
\begin{equation} \label{dimlessJ}
  \bm{J}_{\textrm{s}}\left(\lambda\bm{r}',\displaystyle\frac{\hbar}{\alpha}t'\right)=J_0\bm{J}_{\textrm{s}}'(\bm{r}',t'), \ \bm{J}_{\textrm{n}}\left(\lambda\bm{r}',\displaystyle\frac{\hbar}{\alpha}t'\right)=J_0\bm{J}_{\textrm{n}}'(\bm{r}',t'), \ J_0 = \frac{\sqrt{2}B_{\textrm{c}}}{\mu_0 \lambda}
\end{equation}
\begin{equation}
  \sigma_{\textrm{n}}=\frac{\hbar}{\alpha \mu_0 \lambda^2}\sigma_{\textrm{n}}'
\end{equation}
\begin{equation}
K\left(\lambda\bm{r}',\frac{\hbar}{ \alpha}t'\right)=  K_{\textrm{n}} K'(\bm{r}',t')  
\end{equation}
\begin{equation}
  k = \frac{K_{\textrm{s}} }{K_{\textrm{n}} }
\end{equation}
\begin{equation}
  C = \frac{K_{\textrm{n}}C'}{\lambda^2 \alpha} 
\end{equation} 
We obtain Eq.~\eqref{eqsystems} in the main text. by rewriting the system of equations using these dimensionless quantities. Hereafter, we omit the primes for simplicity.

\subsubsection{Boundary Conditions}
In this section, we summarize the boundary conditions. For the order parameter amplitude $f$, we impose the following boundary condition
\begin{equation} \label{bc_f_Neumann}
  \left[\frac{1}{\kappa}\grad - i\bm{Q}(\bm{r},t)\right]f(\bm{r},t) \cdot \bm{n} = 0
\end{equation}
on all boundaries, where $\bm{n}$ is the outward unit normal vector. This condition follows from minimizing the GL free energy with respect to $f$ at the boundary.

Next, we consider the boundary conditions for the electromagnetic potentials, which are determined by the requirement that no current flows through the boundaries. From Eq.~\eqref{jsQ}, the condition that the supercurrent does not cross the boundaries: $\bm{J}_{\textrm{s}} \cdot \bm{n} = 0$ is given by
\begin{equation} \label{bc_Q}
  \bm{Q}(\bm{r},t) \cdot \bm{n} = 0,
\end{equation}
where $\bm{n}$ is the outward unit normal vector. For the normal current, the requirement $\bm{J}_{\textrm{n}} \cdot \bm{n} = 0$ with Eq.~\eqref{jnPQ} imposes the boundary condition as
\begin{equation} \label{bc_P}
  \left( \grad P(\bm{r},t) + \frac{\partial \bm{Q}(\bm{r},t)}{\partial t} \right) \cdot \bm{n} = 0.
\end{equation}
Note that since Eq.~\eqref{bc_Q} holds at all times on the boundaries, its time derivative $\partial_t \bm{Q} \cdot \bm{n}$ also vanishes. Consequently, Eq.~\eqref{bc_P} is effectively equivalent to $\grad P \cdot \bm{n} = 0$ in the present study.

Finally, we set the boundary conditions for the temperature. At the boundaries in contact with the heat baths, we assume that the temperature of the superconductor is fixed to the bath temperature:
\begin{subequations}
    \begin{align}
    \tau(-L/2,y,t) &= \tau_{\textrm{L}}, \\
    \tau(+L/2,y,t) &= \tau_{\textrm{R}}.
    \end{align}
\end{subequations}
At the vacuum boundaries, the heat flux vanishes. From Eq.~\eqref{SMdefofq}, we obtain
\begin{equation}
  \left. \frac{\partial}{\partial y}\tau(x,y,t) \right|_{y=\pm L/2} = 0.
\end{equation}
The complete set of boundary conditions is summarized as follows:
\begin{subequations} \label{vortexbcsSM}
    \begin{align} 
    \left[\frac{1}{\kappa}\grad - i\bm{Q}(\bm{r},t)\right]f(\bm{r},t) \cdot \bm{n} &= 0,\label{bcsforf_neumann}  \\ 
    \bm{Q}(\bm{r},t) \cdot \bm{n} &= 0, \\
    \grad P(\bm{r},t) \cdot \bm{n} &= 0, \\
    \tau(-L/2,y,t) &= \tau_{\textrm{L}}, \\
    \tau(+L/2,y,t) &= \tau_{\textrm{R}}, \\
    \left. \frac{\partial}{\partial y}\tau(x,y,t) \right|_{y=\pm L/2} &= 0. \label{bctauy}
    \end{align}
\end{subequations}

\subsection{Case with Spin Density Gradient}
In this section, we present the system of equations describing vortex motion under a spin density gradient. We introduce the local spin accumulation $\mu(\bm{r},t)$ in place of the local temperature, assuming that the local transition temperature depends on the square of the spin accumulation. The TDGL equation then takes the following form:
\begin{align}
\label{TDGL2spin}
\MoveEqLeft{\gamma\left(\frac{\partial}{\partial t}+\frac{ie^{*}}{\hbar}\Phi(\bm{r},t)\right)\Psi(\bm{r},t)} \notag
  \\&= \frac{\hbar^2}{2m^{*}}\left(\bm{\bm{\nabla}}-\frac{ie^{*}\bm{A}(\bm{r},t)}{\hbar}\right)^2\Psi(\bm{r},t)-\alpha_0 \frac{T-T_\textrm{c}\left(\mu^2(\bm{r},t)\right)}{T_\textrm{c}\left(\mu^2(\bm{r},t)\right)} \Psi(\bm{r},t)-\beta \left| \Psi(\bm{r},t) \right|^2\Psi(\bm{r},t).
\end{align}
We write the spin diffusion equation for the square of the spin accumulation in dimensional form as
\begin{equation}
  \frac{\partial \mu^2(\bm{r},t)}{\partial t} = \div \left[\sigma^{\textrm{sp}}(\bm{r},t) \grad  \mu^2(\bm{r},t)\right]-\frac{\sigma^{\textrm{sp}}(\bm{r},t)}{2\mu^2(\bm{r},t)}\left(\grad \mu^2(\bm{r},t)\right)^2-\frac{2\mu^2(\bm{r},t)}{\tau^{\textrm{sp}}(\bm{r},t)}.
\end{equation}
The definitions of Ampère's law, normal current density, and supercurrent density remain unchanged. For the non-dimensionalization, we redefine the scale for the order parameter amplitude using
\begin{equation} \label{fbulkspin}
  f_{\textrm{bulk}} = \sqrt{\frac{\alpha_0}{\beta}\frac{T_\textrm{c}(0)-T}{T_\textrm{c}(0)}},
\end{equation}
as $f'=f/ f_{\textrm{bulk}}$, and the spin accumulation as $\mu' = \mu/\mu_{\textrm{R}}$ with the boundary value $\mu_{\textrm{R}}$. Furthermore, we introduce the dimensionless quantities $\sigma'^{\textrm{sp}} = \sigma^{\textrm{sp}} / \sigma^{\textrm{sp}}_{\textrm{n}}$ and $\tau'^{\textrm{sp}} = \tau^{\textrm{sp}} / \tau^{\textrm{sp}}_{\textrm{n}}$, using their respective values in the normal state as characteristic scales. In this dimensionless framework, the spin diffusion equation takes the following form, where we hereafter omit primes for simplicity:
\begin{equation}
a_1 \frac{\partial \mu^2(\bm{r},t)}{\partial t} = \bm{\nabla} \cdot \left[ \sigma^{\textrm{sp}}(\bm{r},t) \bm{\nabla} \mu^2(\bm{r},t) \right] - \frac{\sigma^{\textrm{sp}}(\bm{r},t)}{2\mu^2(\bm{r},t)} \left( \bm{\nabla} \mu^2(\bm{r},t) \right)^2 - a_2 \frac{2\mu^2(\bm{r},t)}{\tau^{\textrm{sp}}(\bm{r},t)},
\end{equation}
where $a_1 = \alpha \lambda^2 / (\hbar \sigma^{\textrm{sp}}_{\textrm{n}})$ and $a_2 = \lambda^2 / (\sigma^{\textrm{sp}}_{\textrm{n} } \tau^{\textrm{sp}}_{\textrm{n}})$ are dimensionless constants. Rewriting the system with the units defined in the previous section, we arrive at Eqs.~\eqref{eqsystemsspin}. 
We choose the boundary conditions for the order parameter as
\begin{subequations} \label{vortexbcsspin}
   \begin{align}
    \left[\left(\frac{1}{\kappa}\grad - i\bm{Q}(\bm{r},t)\right)f(\bm{r},t)\right]\cdot \bm{n}&=0\\
    \bm{Q}(\bm{r},t) \cdot \bm{n} &= 0 \\
    \left( \grad P(\bm{r},t) + \frac{\partial \bm{Q}(\bm{r},t)}{\partial t} \right) \cdot \bm{n} &= 0 \\
    \mu^2(-L/2,y,t) &= 0 \\
    \mu^2(+L/2,y,t) &= \mu^2_{\textrm{R}} \\
    \left. \frac{\partial}{\partial y}\mu^2(x,y,t) \right|_{y=\pm L/2} &= 0. \label{bcmu2ysm}
    \end{align}
\end{subequations}